\documentclass[12pt]{spieman}  
\usepackage{amsmath,amsfonts,amssymb}
\usepackage{graphicx}
\usepackage{setspace}
\usepackage{tocloft}
\usepackage{soul}
\sethlcolor{yellow}
\usepackage[normalem]{ulem}
\title{Deep Learning for Atmospheric Turbulence Mitigation: A Comparison Between GANs and Transformers}

\author[a,*]{Juan S. Carrillo}
\author[a,b]{Jérôme Gilles}
\affil[a]{San Diego State University, Computational Science Research Center, Department of Mathematics and Statistics, 5500 Campanile Drive, San Diego, CA 92182, USA}
\affil[b]{San Diego State University, Department of Mathematics and Statistics, 5500 Campanile Drive, San Diego, CA 92182, USA}

\cftpagenumbersoff{figure}
\cftpagenumbersoff{table} 
\begin{document} 
\maketitle

\begin{abstract}
Atmospheric turbulence creates complex spatial and temporal distortions that significantly reduce image quality. This work compares two deep learning methods for turbulence reduction: Generative Adversarial Networks (GANs) and Transformer-based models. Both methods are trained and tested on the SOTIS dataset, which includes multi-frame sequences under different turbulence conditions with ground truth references. Our results show an important finding: GAN-based models perform better in severely degraded, low-quality situations, while Transformer-based models achieve better results in moderate to high-quality settings, especially in correcting geometric distortions. We also enhance the Turbulence Mitigation Transformer (TMT) by introducing a refined training strategy, an augmented input design that includes a temporally averaged frame as prior structural information, and a modified loss function that incorporates the Structural Similarity Index (SSIM). These improvements lead to consistent performance gains across various turbulence conditions. Overall, this study provides an in-depth evaluation of deep learning methods for turbulence reduction and offers practical guidance for selecting models in different imaging environments.
\end{abstract}

\keywords{image restoration, atmospheric turbulence, GAN, Transformer, deep learning, SSIM}

{\noindent \footnotesize\textbf{*}Juan S. Carrillo,  \linkable{jcarrillo5698@sdsu.edu} }

\begin{spacing}{2}   

\section{Introduction}
\label{sect:intro}  
Earth's atmosphere introduces image distortions due to spatial and temporal variations in temperature and air movement. As light propagates through this turbulent medium, electromagnetic waves are deflected along non-uniform and time-varying paths, resulting in degraded image quality. These distortions—collectively referred to as atmospheric turbulence—manifest as motion blur, geometric warping, and intensity fluctuations due to refractive index variations in the air \cite{su2022restoration}.

Restoring turbulence-degraded imagery is critical for applications requiring high spatial resolution and structural fidelity. Solutions can be broadly categorized into hardware and software approaches. Hardware-based techniques, such as adaptive optics or custom optical systems, offer real-time correction but are expensive and often tailored to specific setups. Software-based methods, by contrast, aim to recover clean imagery using post-processing algorithms applied to turbulence-distorted inputs, making them more flexible and cost-effective.

A commonly used forward model describing the turbulence-degraded image formation process is given in \cite{jain2022evaluation} and expressed in Equation~(\ref{eq:forMod}):
\begin{equation}
\label{eq:forMod}
   f_i = \Phi_i (H \ast u) + n_i
\end{equation}
where $f_i$ is the observed image at time $i$, $u$ is the clean latent image, $\Phi_i$ represents geometric distortions, $H$ is the point spread function (PSF) modeling the blur, and $n_i$ is additive noise.

Classical restoration methods for estimating $u$ include  semi-blind deconvolution \cite{deledalle2020blind} (e.g., maximum likelihood estimation \cite{levin2011efficient}, total variation (L$^1$) regularization\cite{rudin1992nonlinear}) and Wiener filtering\cite{levin2011efficient}. However, while these techniques were widely used in the past, convolutional neural networks (CNNs) began showing superior results even a decade ago in challenging image degradation scenarios such as motion blur and geometric distortion, also referred to as tilt \cite{yuan2022dynamic}. This trend has only intensified in recent years with the development of specialized architectures for turbulence mitigation.

These models have been applied beyond imaging, including in free-space optical (FSO) communication systems. For example, compact CNN-based architectures have been proposed for turbulence compensation at the transmitter, receiver, or jointly across the FSO link \cite{amirabadi2022low}. Convolutional networks have also been used to correct orbital angular momentum (OAM) beam distortions, enhancing mode purity and classification accuracy under varying conditions \cite{liu2019deep}. A comprehensive review of learning-based turbulence mitigation methods—including CNNs, transformer models, and selective state-space approaches—is provided in \cite{hill2025deep}. Recent advances such as temporal–spatial residual Wasserstein Generative Adversarial Networks (TSR-WGAN) leverage spatial–temporal consistency to model residual distortions in image sequences \cite{jin2021neutralizing,rai2022removing} Additionally, Vorontsov et al.\cite{vorontsov2020atmospheric} introduced a deep network trained on intensity scintillation patterns to estimate the refractive index structure parameter $C_n^2$, enabling high-resolution turbulence diagnostics and comparison with theoretical models.

For image deblurring and turbulence restoration, deep learning architectures such as Denoising CNNs (DnCNN), Conditional Generative Adversarial Networks (CGANs), and Fully Connected Networks (FCNs) have also been proposed. In this study, we evaluate and compare two state-of-the-art paradigms—Generative Adversarial Networks and Transformer-based models—for the restoration of turbulence-degraded images.

\subsection{Problem Statement, Scope, Novelty, and Contributions}

Despite the rapid development of deep learning approaches for atmospheric turbulence mitigation, there remains limited understanding of how different architectural paradigms behave under varying turbulence conditions. Existing studies typically focus on the development of a single restoration architecture, making it difficult to assess the relative strengths and limitations of adversarial and Transformer-based approaches within a common experimental framework. Furthermore, the impact of turbulence severity on model selection has not been systematically analyzed, despite its practical importance in real-world imaging applications.

The novelty of this work lies in combining a controlled cross-architecture evaluation with targeted modifications to the TMT training and input formulation. Rather than introducing a new backbone architecture, this study investigates how fundamentally different restoration paradigms behave when evaluated under the same turbulence conditions and identifies specific modifications that improve Transformer-based turbulence restoration.

More specifically, the original contributions of this work are threefold. First, GAN, generator-only, original TMT, modified TMT, and MAMAT approaches are evaluated within a common experimental framework and stratified according to turbulence severity, allowing the performance regimes of the different restoration paradigms to be characterized rather than reporting only aggregate performance. Second, the original TMT training procedure is modified by enforcing turbulence-consistent frame selection and introducing a composite SSIM--Charbonnier--MSE objective to improve structural supervision during optimization. Third, an augmented TMT input representation is introduced in which a temporally averaged frame is appended to the multi-frame sequence as an additional structural prior. Experimental results show that this additional reference consistently improves SSIM across all evaluated bins, with the largest gain observed in the intermediate-quality regime.

Beyond restoration accuracy, the proposed evaluation also examines cross-dataset behavior and computational complexity, providing insight into the generalization and efficiency trade-offs associated with the different model families. Collectively, these contributions distinguish the present work from studies that introduce or evaluate a single turbulence-restoration architecture and provide a systematic analysis of how model design, training strategy, and input representation affect performance across different turbulence regimes.

\section{Methodology}\label{sec:metodologia}

\subsection{Dataset}
One of the main challenges in applying artificial intelligence algorithms, such as neural networks, is the need for high-quality datasets to enable effective training. The performance of such models depends heavily on both the quality and variability of the training data. In this work, we use the SOTIS dataset introduced by Jain et al.~\cite{jain2022evaluation}, which contains a large number of real images affected by simulated atmospheric turbulence, as well as corresponding ground truth images without noise or distortion. This makes it a strong candidate for training and evaluating restoration models.

The dataset simulates various levels of turbulence by modifying two parameters: the refractive index structure constant $C_n^2$ and the propagation distance $L$ (in kilometers) from the sensor. Specifically, the turbulence strength is defined as $C_n^2 = a \cdot 10^{-b}$, with $a = \{1, 3, 5, 7\}$ and $b = \{14, 15, 16, 17\}$.

The SOTIS dataset includes blurred images affected by tilt distortions, along with corresponding undistorted ground truth references. This comprehensive configuration enables robust evaluation of image restoration models across different distortion levels and allows direct comparison against clean reference images.

\begin{figure}
\begin{center}
\begin{tabular}{c}
\includegraphics[height=7.5cm]{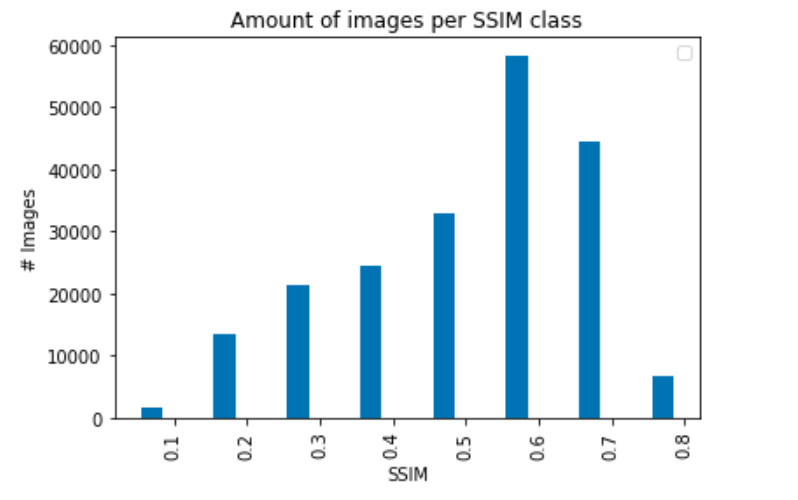}
\end{tabular}
\end{center}
\caption 
{ \label{fig:cantImg}
Distribution of images in the SOTIS dataset across turbulence conditions. } 
\end{figure}

The Structural Similarity Index (SSIM) is a perceptual metric that quantifies image quality degradation, with values closer to 1 indicating higher similarity to the reference image. Figure~\ref{fig:cantImg} shows the distribution of image samples in the SOTIS dataset, stratified by SSIM values. Most of the images fall into intermediate SSIM ranges (0.5–0.7), representing moderate turbulence conditions. This stratification allows for targeted evaluation of model performance across varying distortion levels.

\subsection{Pipeline}

\begin{figure}
\begin{center}
\begin{tabular}{c}
\includegraphics[width=15cm]{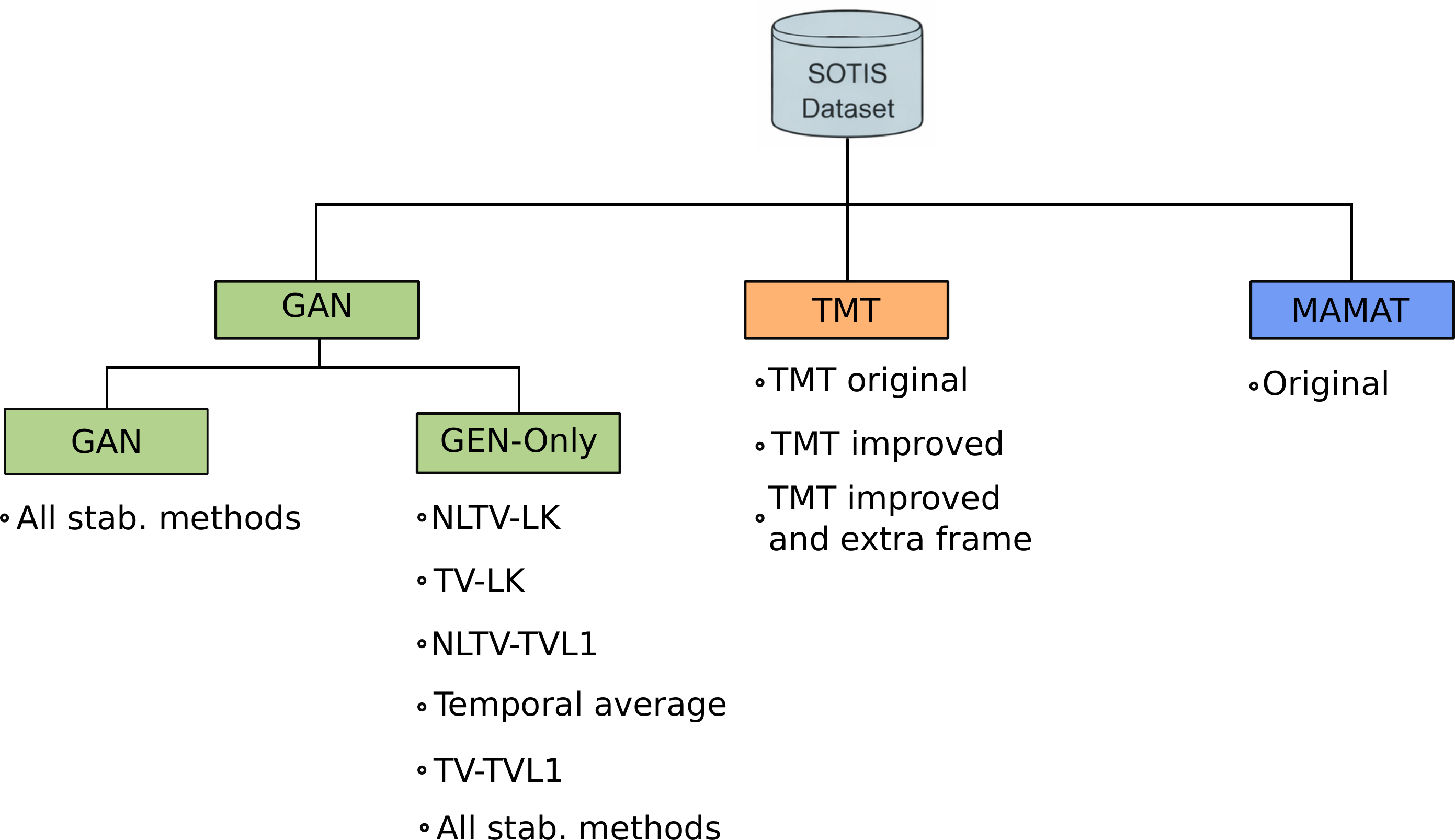}
\end{tabular}
\end{center}
\caption {Overview of the experimental pipeline used in this study. Starting from the SOTIS dataset, three main branches are evaluated: GAN based models, Transformer based TMT models and the MAMAT model included as an external reference method.
} \label{fig:pipeline}
\end{figure} 

To provide a clear and systematic comparison between learning-based approaches for atmospheric turbulence mitigation, this study adopts a structured experimental pipeline that organizes all trained models into distinct architectural families. An overview of the full pipeline, including all model branches and training variants, is illustrated in Fig.~\ref{fig:pipeline}. This section is intended to describe the overall experimental organization; detailed descriptions of each algorithm, architecture, and training procedure are provided in the following sections.

All experiments are conducted using the SOTIS dataset, which serves as a common and controlled source of turbulence-degraded imagery and corresponding ground truth. From this dataset, three independent branches are considered: GAN-based models, Turbulence Mitigation Transformer (TMT) based models, and the 3D Mamba-Based Atmospheric Turbulence Removal (MAMAT) algorithm, which is included as an external reference method.

To ensure a fair comparison, each architecture was independently optimized to achieve its best overall performance before the comparative evaluation was conducted. Although GAN-based and Transformer-based models require different optimization strategies due to their fundamentally different architectures, all models were trained using the same training and validation sets and evaluated on the same test set using identical performance metrics. This common experimental framework ensures that the reported differences reflect the behavior of the restoration paradigms rather than differences in the evaluation protocol.

The generative adversarial network (GAN)-based branch is divided into two independent training paths, reflecting different modeling assumptions and objectives. The first path corresponds to a full GAN model trained using all stabilization methods jointly, where both the generator and discriminator are optimized using stabilized inputs produced by multiple preprocessing techniques. This configuration evaluates the GAN’s ability to generalize across a diverse set of stabilized representations.

The second path corresponds to a generator-only (GEN-only) configuration, in which only the generator network is trained. In this branch, separate GEN models are trained using stabilized inputs produced by individual stabilization methods, including Non-Local Total Variation-Lucas-Kanade (NLTV--LK), Non-Local Total Variation-TVL1 (NLTV--TVL1), Total Variation--Lucas--Kanade (TV--LK), Total Variation--TVL1 (TV--TVL1), and temporal averaging, as well as an additional GEN model trained using all stabilization methods combined. These GEN variants are trained independently and are not connected to the full GAN configuration. This separation allows the influence of each stabilization strategy to be assessed independently of adversarial training.

The Transformer-based branch consists of three variants derived from the TMT architecture. These include the original TMT model, improved architectural variants introduced in this work, and an enhanced version that incorporates a temporally averaged frame as additional input. Unlike the GAN-based models, TMT operates directly on multi-frame sequences and is trained end-to-end to jointly address geometric distortions without external stabilization preprocessing.

In addition to the learning-based models trained in this study, the MAMAT algorithm ~\cite{hill2025mamat} is included as a reference method. The MAMAT model is trained directly from the SOTIS dataset and forms an independent branch in the experimental pipeline. It is not modified or extended in this work and is used solely as an external comparison baseline rather than a central focus of the analysis.

Similarly, the original TMT architecture, as introduced in prior work, is evaluated as an external reference within the Transformer-based branch. While it shares the same dataset and evaluation protocol, it serves as a baseline against which the improved and enhanced TMT variants proposed in this study are assessed.

\subsection{Generative Adversarial Network model}

\begin{figure}
\begin{center}
\begin{tabular}{c}
\includegraphics[height=5.5cm]{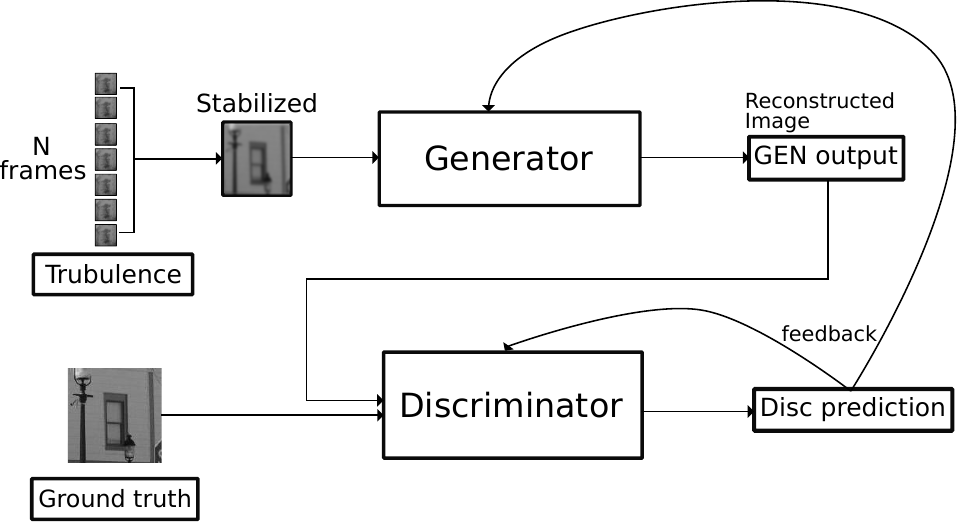}
\end{tabular}
\end{center}
\caption{Overview of the GAN based restoration pipeline. Turbulence degraded image sequences are first stabilized and provided as input to the generator, which produces a reconstructed image. \label{fig:GANarq}
}
\end{figure}

Generative Adversarial Networks have shown promising results in image generation and image restoration tasks. A GAN is composed of two neural networks, a generator and a discriminator, which are trained simultaneously through an adversarial learning process. As illustrated in Fig.~\ref{fig:GANarq}, stabilized inputs are processed by the generator to produce a reconstructed image, while the discriminator compares the generated output with the ground truth image and outputs a prediction that guides the training process. Over the course of training, the generator progressively improves its ability to produce realistic reconstructions that are increasingly difficult for the discriminator to distinguish from real images.
 ~\cite{kupyn2019deblurgan}.

\begin{figure}
\begin{center}
\begin{tabular}{c}
\includegraphics[width=16cm]{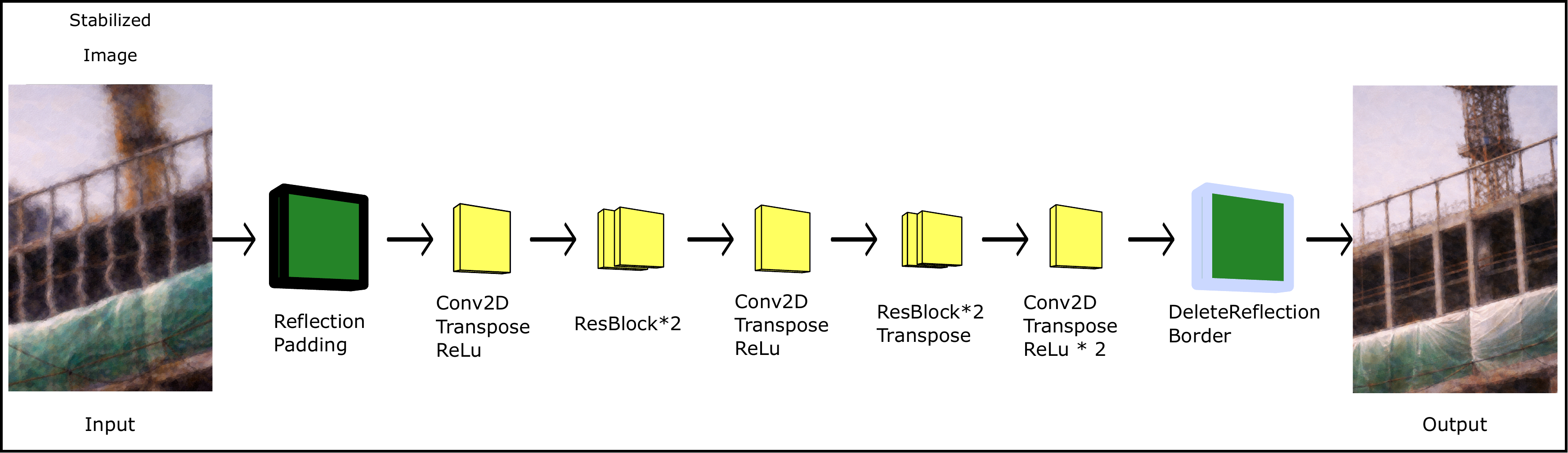}
\end{tabular}
\end{center}
\caption 
{ \label{fig:arqui}
Generator architecture used in the GAN model. It includes reflected edge padding, transposed convolutions, and residual blocks for stable training. } 
\end{figure}

Figure~\ref{fig:arqui} shows the architecture of the proposed generator. It begins with a padding layer that adds a reflected edge to avoid boundary artifacts. This padding is removed at the output to preserve the original image dimensions. The hidden layers use transposed convolutions, which help maintain the image size without resorting to pooling operations that could lead to information loss. Additionally, ResBlock-style ~\cite{he2016deep} convolutional blocks are employed to facilitate stable training and improve convergence.

\begin{figure}
\begin{center}
\begin{tabular}{c}
\includegraphics[height=4.4cm]{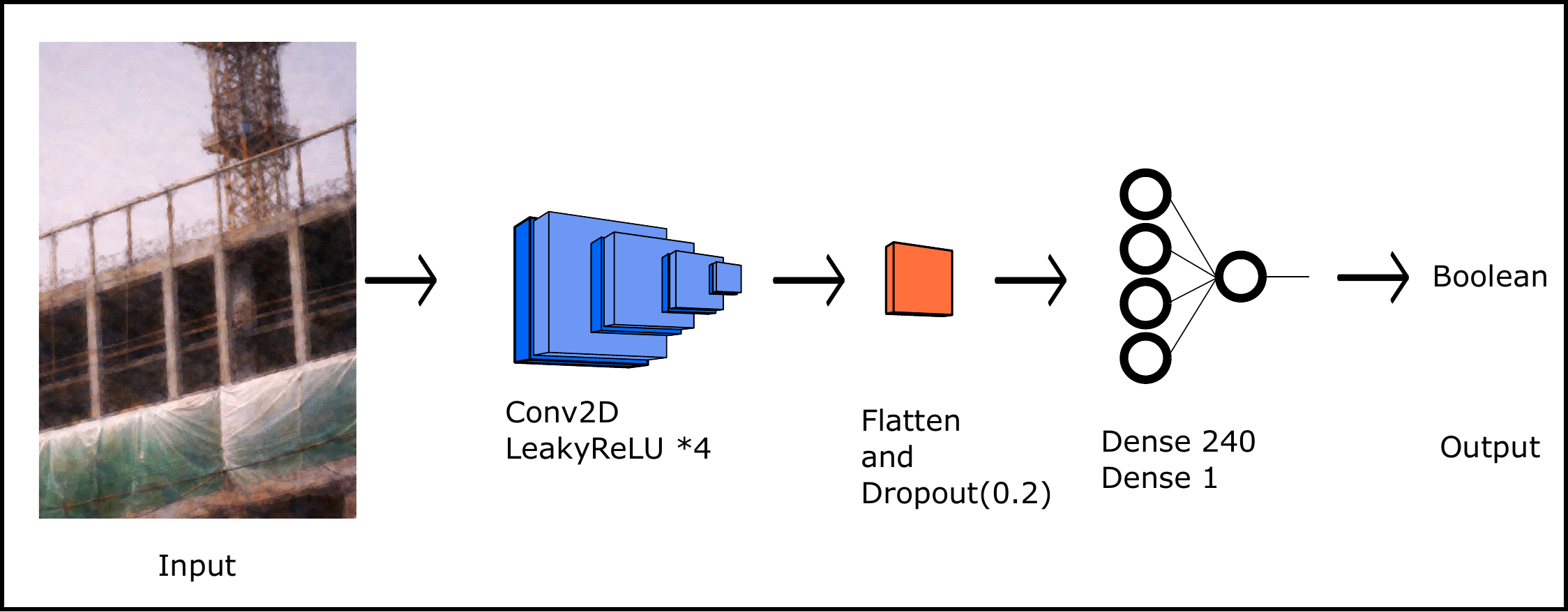}
\end{tabular}
\end{center}
\caption 
{ \label{fig:arquiDisc}
Discriminator architecture composed of convolutional layers and a fully connected classifier that distinguishes real (clean) from generated images. } 
\end{figure}

The architecture of the discriminator is shown in Fig.~\ref{fig:arquiDisc}. It consists of four 2D convolutional layers, followed by resizing, flattening, and dropout layers. A final fully connected layer outputs a binary prediction indicating whether the input image is classified as \emph{real} (free of geometric distortion and blur) or \emph{fake} (turbulence-degraded). This real-versus-fake formulation follows the standard GAN paradigm, adapted here to distinguish clean, turbulence-free images from restored or turbulent ones. The discriminator is trained to classify ground truth images as real and all reconstructed or turbulence-degraded images as fake, providing adversarial feedback that guides the optimization of the generator.

\subsection{Pre-processing images for GAN}
The GAN architecture implemented in this study operates on single-frame inputs and therefore lacks any mechanism to handle geometric distortions or temporal alignment across frames. As a result, the model is primarily effective for learning to correct blur and photometric degradation, but not for compensating spatial misalignment such as tilt or warping. To mitigate this limitation, all turbulence-distorted images were first preprocessed using stabilization algorithms that reduce geometric distortions before being passed to the GAN. This preprocessing step ensures that the generator receives inputs that are spatially aligned with their corresponding ground truth targets, allowing the model to focus on restoring sharpness and structural detail rather than learning alignment from scratch.

To improve generalization and robustness, the GAN was trained using stabilized images produced by five distinct stabilization methods: NLTV–LK, NLTV–TVL1, TV–LK, TV–TVL1, and a Temporal Average (T.A.) baseline. Each method combines a specific \textit{regularization strategy} with an \textit{optical flow estimation algorithm}. The abbreviations correspond to:  
\begin{itemize}
    \item \textbf{TV}: Total Variation regularization, which enforces spatial smoothness while preserving image edges~\cite{rudin1992nonlinear}.  
    \item \textbf{NLTV}: Nonlocal Total Variation regularization, an extension of TV that incorporates nonlocal pixel similarities, improving performance in textured or repetitive regions~\cite{gilboa2009nonlocal}.  
    \item \textbf{LK}: Lucas–Kanade optical flow, a fast local method that estimates pixel displacements based on intensity gradients~\cite{bouguet2001pyramidal}.  
    \item \textbf{TV-L1}: Total Variation-$L_1$ optical flow, a robust variational approach that minimizes an $L_1$ data term combined with TV-based smoothness, offering improved performance under noise and large displacements~\cite{sanchezpreprint}.  
    \item \textbf{Temporal Average (TA)}: a simple frame-wise averaging technique that reduces random turbulence but does not explicitly correct geometric distortions.  
\end{itemize}

Each stabilization approach compensates turbulence induced motion differently, introducing unique residual artifacts. In general, these methods combine two complementary components: a regularization term that constrains the reconstructed image to remain spatially coherent, and an optical flow algorithm that estimates local geometric deformations between frames, aligning them to a common reference. As described in the Mao-Gilles algorithm for turbulence stabilization~\cite{gilles2013mao}, this coupling between optical flow and regularization ensures stable convergence and visually consistent results.

Although the five stabilization approaches differ in their mathematical formulation, their combined use during GAN training exposes the model to a broad range of stabilized inputs and residual distortions. This diversity encourages the GAN to learn invariant restoration features and reduces overfitting to the artifacts of any single pre-processing method. In practice, GAN training is known to be sensitive to instability and convergence issues. For this reason, instead of selecting a single stabilization method that may yield slightly higher reconstruction quality in isolated cases, all stabilization folders were used jointly during training to promote greater robustness and improve convergence behavior.

This decision prioritizes stability and generalization over maximizing performance for a specific stabilization strategy. The main GAN section and its corresponding results therefore focus on the model trained with all stabilization methods combined, rather than exploring separate adversarial training for each individual stabilization method. A detailed comparison of individual stabilization strategies is instead performed in the generator-only configuration, where training stability is less critical and the effect of each stabilization method can be analyzed more directly.

\subsection{Training the GAN Model}

Training the GAN model involves two stages: first training the discriminator independently, and then training the generator as part of the full adversarial loop. During the initial stage, the discriminator is trained to correctly classify:
\begin{enumerate}
  \item Stabilized images as class~0,
  \item Ground truth images without turbulence as class~1,
  \item Images generated by the generator as class~0. The generator is initially randomized, so its outputs do not yet resemble the ground truth images.
\end{enumerate}

This approach allows the discriminator to learn to differentiate between real and synthetic images while also reinforcing its ability to reject distorted inputs. The generator, in turn, improves by learning to produce images that can fool the discriminator.

In the second stage, the complete GAN is trained. The discriminator’s weights are frozen (non-trainable), and only the generator is updated. The generator takes stabilized distorted images as input and is trained to produce outputs that the discriminator classifies as class~1, i.e., indistinguishable from clean, turbulence-free images. By aiming for class~1, the generator attempts to fool the discriminator into believing that its synthetic outputs are real undistorted images.

The training setup uses the Adam optimizer for both networks. For the discriminator, the learning rate was set to $9.5 \times 10^{-6}$, while the generator was trained with a learning rate of $3 \times 10^{-4}$. Binary cross-entropy was used as the loss function for both networks. The model was trained using a batch size of 16 for 12~epochs, with a total of 208{,}000 images from the SOTIS dataset. The relatively low number of epochs reflects the large size and diversity of the training set. A similar training configuration, combining binary cross-entropy and the Adam optimizer, has been successfully applied in related deblurring tasks, notably in DeblurGAN~\cite{kupyn2018deblurgan}, demonstrating the effectiveness of adversarial training for blind motion deblurring.

\subsection{Generator-only model}\label{sec:GEN-only}

In addition to the full GAN configuration, we consider a generator-only (GEN) training strategy as a methodological variant of the GAN-based approach. The purpose of this variant is to isolate the contribution of the generator network and evaluate its performance when trained without adversarial feedback. This configuration follows directly from the GAN branch introduced in the experimental pipeline in Fig.~\ref{fig:pipeline} and is intended to address some of the practical limitations observed in adversarial training.

The generator architecture used in the GEN configuration is identical to the generator employed in the full GAN model , showed in Fig.~\ref{fig:arqui}. However, instead of being trained within an adversarial framework, the generator is optimized directly against the ground truth images using a supervised loss function. In this setting, the discriminator is entirely removed, and the generator is trained as a standalone image-to-image restoration network. This design choice ensures that any observed performance differences between the GAN and GEN configurations can be attributed to the presence or absence of adversarial learning rather than architectural changes.

The generator-only model is trained using a combined loss function based on Mean Squared Error (MSE) and the Structural Similarity Index (SSIM). The MSE component encourages pixel wise fidelity to the ground truth, while the SSIM term promotes preservation of structural and perceptual features. The combined loss is defined as:

\begin{equation}
\mathcal{L}_{\text{GEN}}(x, y) = \alpha \, \mathcal{L}_{\text{MSE}}(x, y) + \left(1 - \text{SSIM}(x, y)\right),
\end{equation}
where \( x \) denotes the predicted image, \( y \) the ground truth image, and \( \alpha \) is a weighting parameter that controls the relative contribution of the MSE term. In our experiments, \( \alpha = 0.65 \). This formulation provides a stable optimization objective that balances pixel level accuracy and structural similarity, while avoiding the convergence instabilities commonly associated with adversarial training.

As with the full GAN model, the GEN configuration operates on stabilized single-frame inputs. Atmospheric turbulence introduces significant geometric distortions that cannot be handled by a single-frame generator alone. Consequently, all input images are preprocessed using stabilization techniques prior to training. The stabilization methods employed are the same as those introduced in the GAN section and are discussed in detail in the previous subsection. In this section, stabilization is treated as a preprocessing step that enables meaningful supervised training of the generator by ensuring spatial alignment between inputs and ground truth targets.

Multiple GEN models are trained to assess the impact of different stabilization strategies. Specifically, one independent GEN model is trained for each stabilization method, using the same generator architecture and loss function, as well as an additional GEN model trained using all stabilization methods combined. This design allows the influence of individual stabilization techniques to be evaluated independently while also assessing the benefits of exposing the generator to a broader distribution of stabilized inputs during training.

The removal of the discriminator also introduces important limitations. Without adversarial feedback, the generator no longer benefits from the additional supervisory signal provided by the discriminator during training. Consequently, the model relies entirely on the selected reconstruction losses and the available training data to guide optimization. In contrast, adversarial learning can encourage the generator to capture characteristics of the underlying distribution of clean images that are not explicitly represented by reconstruction losses such as MSE or SSIM alone. While this additional source of information can potentially improve restoration quality, GAN training is also known to be sensitive to imbalances between the generator and discriminator. In practice, the discriminator may converge more rapidly than the generator, reducing the effectiveness of the adversarial feedback and making optimization more difficult. The GEN-only configuration therefore represents a trade-off: it sacrifices the distribution-level information provided by adversarial learning in exchange for improved stability, reduced computational complexity, and more predictable convergence.

\subsection{Transformer Network}\label{sec:TMT}

Transformer-based architectures have emerged as powerful tools for complex image restoration tasks, particularly in scenarios involving spatial and temporal distortions such as those caused by atmospheric turbulence. In this work, we adopt the TMT model introduced by Zhang et al.~\cite{zhang2024imaging}, which is a multi-frame transformer specifically designed for turbulence correction in image sequences.

The TMT design is derived directly from the physical forward model introduced in Equation~(\ref{eq:forMod}), where atmospheric turbulence is modeled as a geometric distortion operator $\Phi_i$ followed by convolution with a blur kernel $H$. In this formulation, image degradation can be interpreted as the successive application of geometric displacement and spatially varying blur to the clean latent image $u$, before additive noise is introduced. 

This decomposition reflects the physical interpretation of turbulence, where low order aberrations induce pixel displacement and higher order aberrations produce blur. By mirroring this structure, TMT separates restoration into a removal distortion stage that compensates for $\Phi_i$ and a deblurring stage that addresses the effect of $H$, enabling the model to learn geometric alignment and fine detail restoration in a physically consistent and interpretable manner.

This architecture addresses three key limitations of prior deep learning models for turbulence mitigation: poor temporal alignment, lack of multi-scale feature aggregation, and insufficient spatial-temporal context modeling. The TMT model introduces components such as temporal-channel joint attention, hierarchical supervision, and a transformer-based encoder-decoder structure.

\subsubsection{Multi-Frame Attention Mechanism}

\begin{figure}
\begin{center}
\begin{tabular}{c}
\includegraphics[width=16cm]{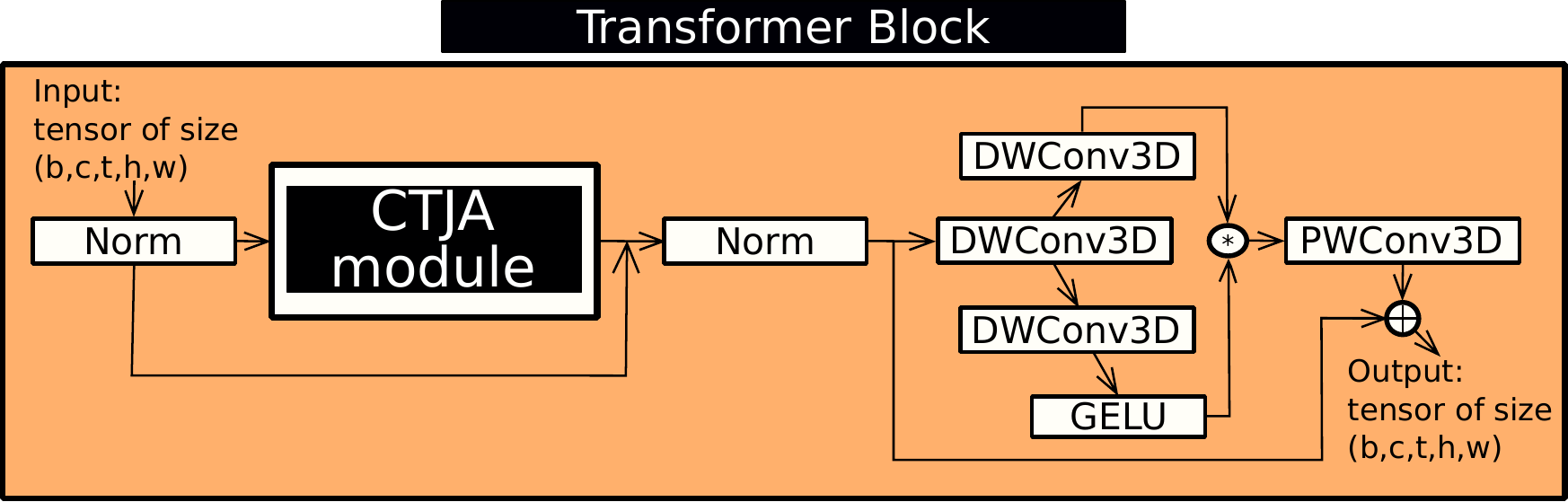}
\end{tabular}
\end{center}
\caption{
\label{fig:trnasfModule}
Internal structure of the Transformer module. It integrates normalization, depth-wise 3D convolutions, GELU activation, residual connections, and the CTJA module shown in Fig.~\ref{fig:CTJA}.
}
\end{figure}

The Channel--Temporal Joint Attention (CTJA) module is the core component responsible for jointly modeling temporal and spatial information across the input sequence. As illustrated in Fig.~\ref{fig:trnasfModule}, each transformer block incorporates CTJA to capture long-range dependencies between consecutive frames while preserving local spatial information.

\begin{figure}
\begin{center}
\begin{tabular}{c}
\includegraphics[width=16cm]{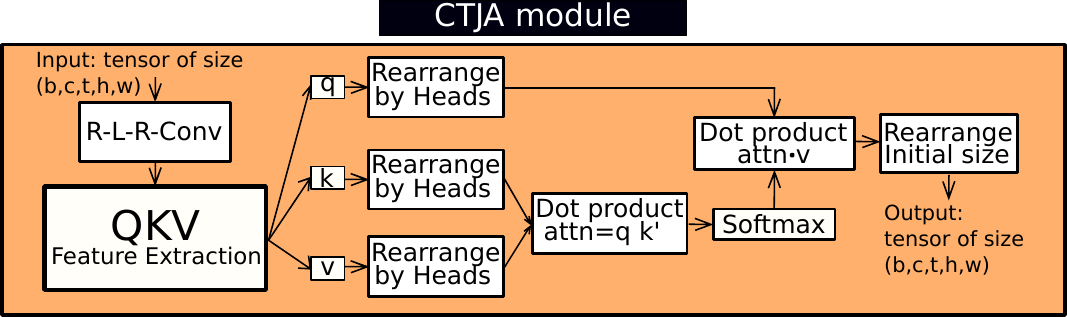}
\end{tabular}
\end{center}
\caption{
\label{fig:CTJA}
Overview of the CTJA module. Queries, keys, and values are extracted after temporal-channel feature mixing and subsequently processed through multi-head attention.
}
\end{figure}

The CTJA module operates on an input tensor
\[
X\in\mathbb{R}^{b\times c\times t\times h\times w},
\]
where $b$ is the batch size, $c$ the number of channels, $t$ the number of frames, and $h\times w$ the spatial resolution. Prior to computing attention, the module applies the rearrange--linear--rearrange and convolution (R-L-R-Conv) transformation proposed in~\cite{zhang2024imaging}, allowing temporal and channel information to interact before the attention operation.

\begin{figure}
\begin{center}
\begin{tabular}{c}
\includegraphics[height=7.5cm]{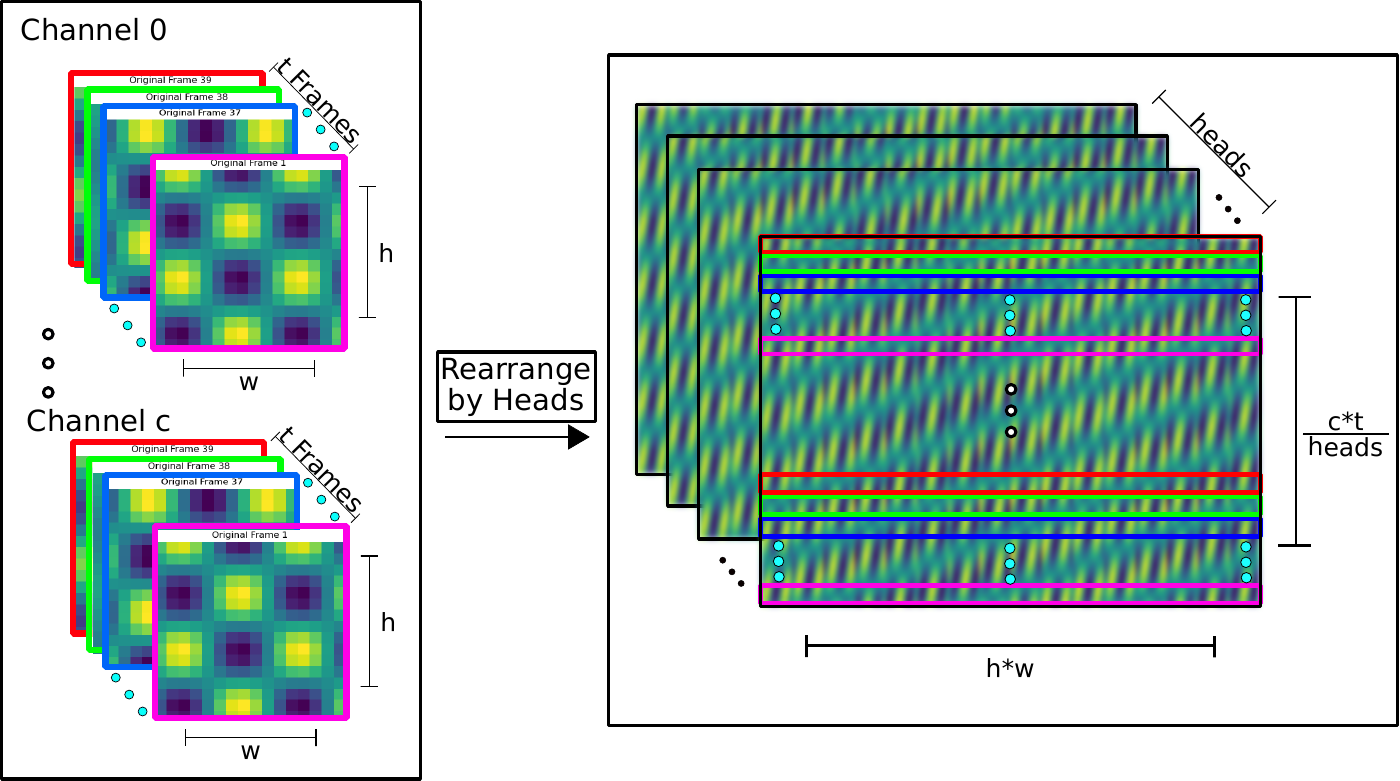}
\end{tabular}
\end{center}
\caption{
\label{fig:rearrangeByHead}
Rearrangement of temporal-channel features into multiple attention heads prior to multi-head attention.
}
\end{figure}

Queries ($Q$), keys ($K$), and values ($V$) are then extracted through learnable linear projections, and attention is computed as

\begin{equation}
\label{eq:attention}
\mathrm{Attention}(Q,K,V)
=
\mathrm{Softmax}
\left(
\frac{QK^\top}{\sqrt{d_k}}
\right)V,
\end{equation}

where $d_k$ denotes the key dimensionality. The resulting feature representation is subsequently divided into multiple attention heads, as illustrated in Fig.~\ref{fig:rearrangeByHead}, enabling different heads to capture complementary temporal and spatial correlations. This mechanism improves temporal coherence and enhances the restoration of turbulence-induced geometric distortions.

\subsubsection{Multi-Scale Supervision}

\begin{figure}
\begin{center}
\begin{tabular}{c}
\includegraphics[width=16cm]{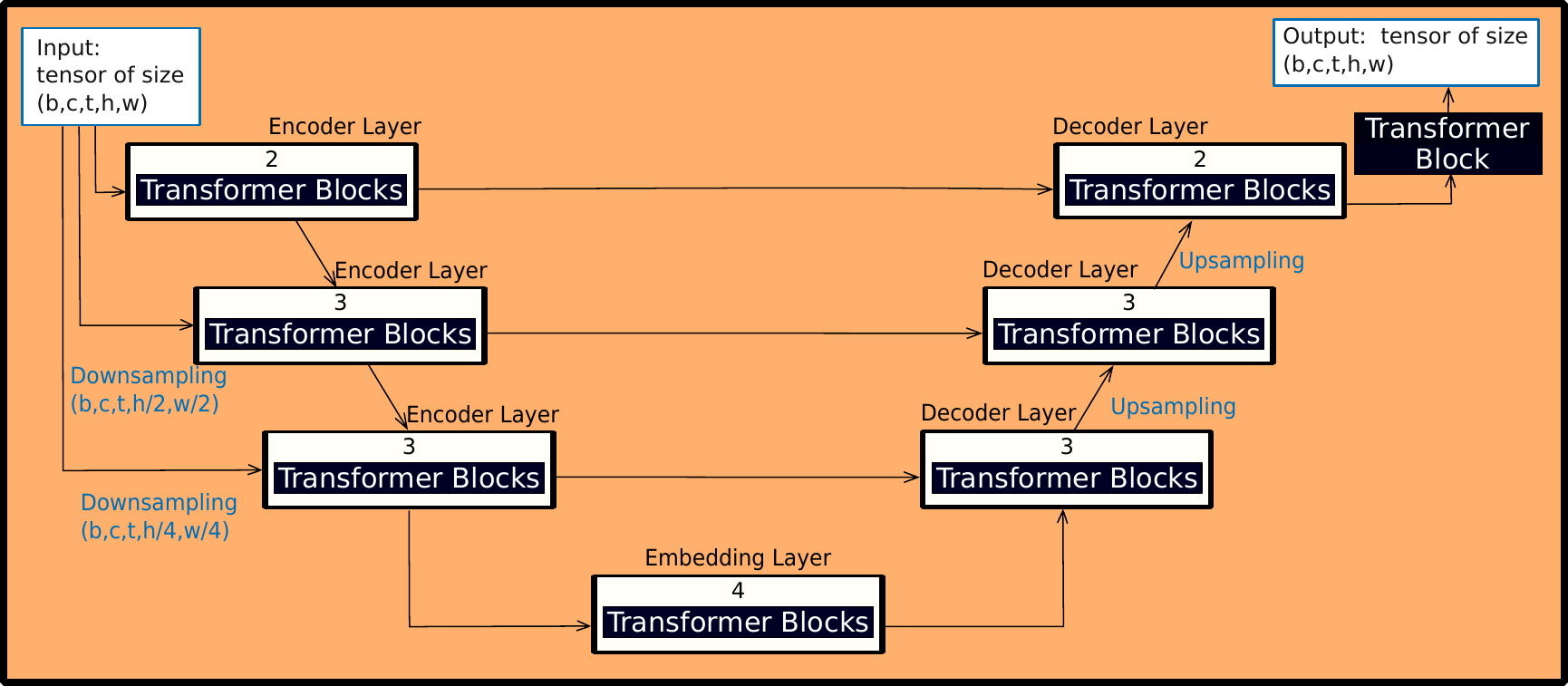}
\end{tabular}
\end{center}
\caption 
{ \label{fig:arqTMT}
Overview of the TMT architecture, which processes a tensor (b, c, t, h, w) through downsampling, transformer-based encoding, and upsampling to reconstruct the input resolution. } 
\end{figure} 
 
The multi-scale structure of the TMT model introduces hierarchical supervision during training. This enables the network to learn features across multiple spatial resolutions, enhancing its ability to address spatially varying distortions. As illustrated in Fig.~\ref{fig:arqTMT}, the input tensor is progressively distributed across multiple layers through downsampling operations, forming a hierarchical representation of spatial features. This structure allows the network to capture both fine- and coarse-scale distortions at different depths. Transformers are well suited for this process due to their capacity to model hierarchical token representations. By incorporating multi-scale loss and supervision, the model generalizes more effectively to real-world turbulence conditions, where distortions may occur at different spatial frequencies. The following subsection further describes how this hierarchical design is implemented within the overall encoder–decoder architecture.

\subsubsection{Encoder–Decoder Architecture with Transformers}

As shown in Fig.~\ref{fig:arqTMT}, the TMT model adopts an encoder–decoder structure in which transformer modules are integrated at each stage of the hierarchical feature pipeline. The input tensor, initially of size $(b, c, t, h, w)$, is progressively downsampled by factors of two, producing intermediate representations such as $(b, c, t, h/2, w/2)$ and $(b, c, t, h/4, w/4)$. These progressively compressed scales allow the encoder to extract spatial and temporal features of increasing abstraction while reducing spatial redundancy. Within each level, transformer blocks operate over these feature maps to jointly capture long-range dependencies across frames and channels.

This expansion and contraction process forms the core of the TMT architecture. During encoding, downsampling aggregates contextual information and enhances the model’s ability to estimate global phase distortions. During decoding, upsampling restores fine spatial details and reconstructs the original resolution $(b, c, t, h, w)$. By fusing information across scales, the model effectively mitigates tilt phase distortions, which are a primary manifestation of turbulence that produce geometric displacements, while recovering sharp structural details. This hierarchical encoder–decoder design balances global context modeling with localized refinement, enabling robust compensation of turbulence-induced motion and blur.

Since the focus of this work is the comparative evaluation and enhancement of the original TMT framework, only the principal architectural components are summarized here; a detailed description of the complete architecture is available in Zhang et al.~\cite{zhang2024imaging}.

\subsection{Training the Transformer Model}

In this study, three Transformer based configurations derived from the Turbulence Mitigation Transformer (TMT) architecture are considered. The first corresponds to the original TMT model as introduced in Section ~\ref{sec:TMT}. The second corresponds to a modified TMT model trained using an improved frame selection strategy designed to enforce temporal consistency in turbulence conditions, together with a refined loss function that improves structural fidelity and training stability. The third configuration builds upon this modified training strategy and additionally incorporates an extra temporally averaged frame as part of the network input. These three configurations allow the impact of each training and input level modification to be evaluated independently.

The TMT models are trained using the SOTIS dataset, which contains image sequences simulated under varying levels of atmospheric turbulence. For the modified training configurations introduced in this work, each input sample consists of a group of frames selected under identical turbulence parameters, characterized by the same values of $C_n^2$ for turbulence strength and $L$ for sensor to target distance. This design choice reflects a realistic physical assumption, namely that in real world imaging scenarios atmospheric turbulence does not vary abruptly over a small number of consecutive frames. By grouping frames with similar turbulence characteristics, the model is encouraged to learn coherent spatiotemporal representations that better reflect realistic acquisition conditions.

In contrast, in the original TMT training formulation, input frames are sampled independently from the same scene without explicitly enforcing similarity in turbulence strength across frames. While this strategy increases data diversity, it may introduce artificial temporal inconsistencies that are unlikely to occur in practice. The modified training strategy adopted here reduces this effect by ensuring that all frames within a sequence share comparable turbulence statistics.

In the third configuration, the modified TMT model further extends the input representation by incorporating an additional temporally averaged frame. Specifically, twelve frames are first selected from the dataset following the turbulence consistency criteria described previously. From this set of frames, an additional frame is computed as the pixel wise temporal average of the selected inputs and appended to the input sequence provided to the network.

This averaged frame serves as a structurally stable reference that captures the dominant spatial content of the scene while suppressing random frame to frame fluctuations induced by atmospheric turbulence. By providing the model with this additional reference, the network is guided toward more consistent structural reconstruction, particularly in regions affected by residual geometric jitter or intensity variations. The impact of including this averaged frame on restoration performance is analyzed separately in Section~\ref{sec:tempuralAverFrame}.

Together, these three configurations enable a structured evaluation of the original TMT baseline, the effect of enforcing temporal consistency during training, and the additional benefit provided by incorporating an averaged reference frame, while maintaining full compatibility with the original TMT architecture.

\subsection{Loss Function}

The choice of loss function plays a critical role in training deep learning models for image restoration, as it directly influences the trade off between perceptual quality, structural fidelity, and pixel wise accuracy. An effective loss function should not only guide the model toward producing outputs that are numerically close to the ground truth, but also ensure that these outputs are visually convincing and free of perceptual artifacts. Relying on a single loss term is often insufficient. Pixel based metrics such as Mean Squared Error (MSE) can encourage accurate reconstructions in terms of intensity values but tend to produce overly smooth results, while perceptual or structure aware metrics like the Structural Similarity Index (SSIM) capture visual quality but may overlook fine pixel level differences \cite{zhao2015loss}. Robust alternatives such as the Charbonnier loss help address outliers and preserve important image details.

In the original TMT formulation, training relied primarily on pixel based loss functions and did not explicitly incorporate SSIM based supervision. In this work, a refined composite loss function is introduced to improve the structural fidelity of the restored images. By explicitly integrating an SSIM based term together with MSE and Charbonnier losses, the training process is encouraged to preserve both local intensity accuracy and global structural consistency.

For the modified TMT models considered in this study, a composite loss function is therefore adopted to combine the strengths of these complementary objectives. This refined loss formulation is designed to improve upon the original TMT training strategy by better aligning numerical accuracy with perceptual and structural quality. The following subsections describe each loss component in detail.

\subsubsection{Structural Similarity Index (SSIM) Loss}

The SSIM loss measures the perceptual similarity between the restored and ground truth images. It is particularly effective at preserving structural information and reducing perceptual artifacts in the output. The Structural Similarity Index (SSIM) takes values in the range \([0, 1]\), where 1 indicates perfect structural similarity and 0 indicates no similarity. Defining the loss as \(1 - \text{SSIM}\) ensures that the loss is minimized when the predicted image is structurally identical to the ground truth.

The loss based on SSIM is given in Equation~(\ref{eq:SSIMeq}) as
\begin{equation}
\label{eq:SSIMeq}
\mathcal{L}_{\text{SSIM}}(x, y) = 1 - \text{SSIM}(x, y),
\end{equation}
where \( x \) is the predicted image and \( y \) is the ground truth image. A lower SSIM loss corresponds to higher structural similarity between the two.

\subsubsection{Mean Squared Error (MSE) Loss}
Equation~(\ref{eq:MSE}) defines the Mean Squared Error (MSE) loss, which penalizes the squared differences between predicted and ground truth pixel intensities and encourages the model to minimize pixel-wise reconstruction error:

\begin{equation}
\label{eq:MSE}
\mathcal{L}_{\text{MSE}}(x, y) = \frac{1}{p} \sum_{i=1}^p (x_i - y_i)^2,
\end{equation}
where \( x \) is the predicted image, \( y \) is the ground truth image, and \( p \) is the total number of pixels.

\subsubsection{Charbonnier Loss}

The Charbonnier loss is a smooth approximation of the $L^1$ loss and is more robust to outliers. It is defined in Eq~(\ref{eq:charb}) as

\begin{equation}
\label{eq:charb}
\mathcal{L}_{\text{Charb}}(x, y) = \frac{1}{p} \sum_{i=1}^p \sqrt{(x_i - y_i)^2 + \epsilon^2},
\end{equation}
where $x$ and $y$ are respectively the predicted and ground truth images, \( p \) is the total number of pixels and \( \epsilon \) is a small constant added for numerical stability, typically $\epsilon = 10^{-3}$.

\subsubsection{Combined Loss}

The total loss function is a weighted sum of SSIM, Charbonnier, and MSE losses, as defined in Eq~(\ref{eq:mixloss}):

\begin{equation}
\label{eq:mixloss}
\mathcal{L}_{\text{Total}} = 0.2 \cdot \mathcal{L}_{\text{SSIM}} + 0.6 \cdot \mathcal{L}_{\text{Charb}} + 0.2 \cdot \mathcal{L}_{\text{MSE}}.
\end{equation}
This specific weighting prioritizes Charbonnier loss (0.6), which acts as a robust pixel-wise loss with improved tolerance to outliers compared to standard $L^1$ or $L^2$ losses. By assigning a lower weight to MSE (0.2), the model avoids over-penalizing small pixel errors, which can lead to overly smooth outputs. The term based on the Structural Similarity Index (SSIM) (0.2), while perceptually motivated, tends to be less sensitive to fine details during training. Including it with a moderate weight ensures structural consistency in the output. This combination strikes a balance: Charbonnier ensures pixel accuracy and robustness, SSIM promotes perceptual structure preservation, and MSE stabilizes convergence. Empirically, similar hybrid weighting schemes have demonstrated improved performance in restoration tasks under real-world distortion scenarios~\cite{zhao2016ssim}.

Although the three loss terms have different mathematical definitions, they are implemented as averaged quantities rather than unnormalized sums. Specifically, the MSE loss is computed using the mean reduction provided by PyTorch, the Charbonnier loss is averaged over all pixels, and the SSIM contribution is expressed as $1-\mathrm{SSIM}$. Consequently, the numerical magnitudes of the three losses remain comparable during optimization. Therefore, the weighting coefficients in Eq.~(\ref{eq:mixloss}) should be interpreted as empirical balancing factors rather than direct indicators of the absolute contribution of each loss component.

Moreover, although the weighting coefficients remain fixed throughout the entire training process, the relative contribution of each loss term is not constant because the numerical value of each component evolves differently during optimization. When structural differences between the reconstructed and ground-truth images are large, the SSIM component typically contributes more strongly because the structural similarity remains low. As the reconstruction quality improves and the SSIM approaches one, the relative influence of the Charbonnier and MSE terms naturally becomes more significant, promoting finer pixel-level refinements while preserving the overall structural consistency of the reconstructed image. Consequently, the effective contribution of each loss component evolves dynamically during optimization, even though the weighting coefficients remain unchanged.

\subsection{Optimization and Learning Rate Scheduling}

The TMT model is trained using the Adam optimizer, which provides efficient handling of sparse gradients and adaptive learning rate updates. Adam combines the advantages of both AdaGrad\cite{duchi2011adaptive} and RMSProp\cite{nguyen2020momentumrnn}, enabling stable and fast convergence during training.
To further improve training stability and convergence speed, a two-stage learning rate scheduling strategy is employed:
\begin{itemize}
    \item \textbf{Gradual Warmup Scheduler:} Introduces a slow increase in learning rate at the beginning of training to prevent instability due to large initial updates.
    \item \textbf{Cosine Annealing with Warm Restarts:} Periodically decays and resets the learning rate following a cosine schedule, which helps the optimizer escape local minima and maintain long-term learning efficiency.
\end{itemize}

This scheduling strategy has demonstrated improved convergence speed and generalization in recent studies \cite{zhang2024wuc}.

\subsection{Enhancing Restoration Using the Temporal Average Frame}\label{sec:tempuralAverFrame}

We propose an enhancement to the TMT input structure by incorporating a temporally averaged frame. The average of multiple distorted frames can provide a reliable estimate of the true underlying structure of the scene. By including this averaged frame as part of the network input, the model receives an additional spatial reference that can help guide the restoration process \cite{gilles2008atmospheric}.

\begin{figure}
\begin{center}
\begin{tabular}{c}
\includegraphics[height=5.5cm]{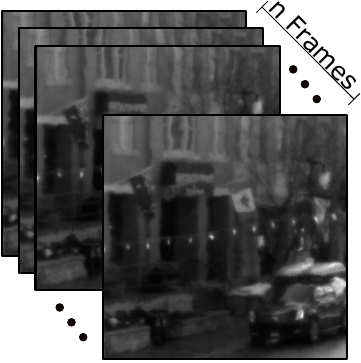}
\end{tabular}
\end{center}
\caption{ \label{fig:nframes}
Temporal sequence of $n$ frames capturing the same scene at different times.}
\end{figure}

Figure~\ref{fig:nframes} illustrates a sequence of $n$ frames capturing the same scene at different time steps under atmospheric turbulence.

To test this hypothesis, the data loading pipeline was modified. Instead of using only the original $n$ frames as input, a new $(n+1)^\text{th}$ frame is computed as the temporal average of the $n$ frames. This averaged frame is appended to the input stack and serves as a stable visual reference.

\begin{figure}
\begin{center}
\begin{tabular}{c}
\includegraphics[width=16cm]{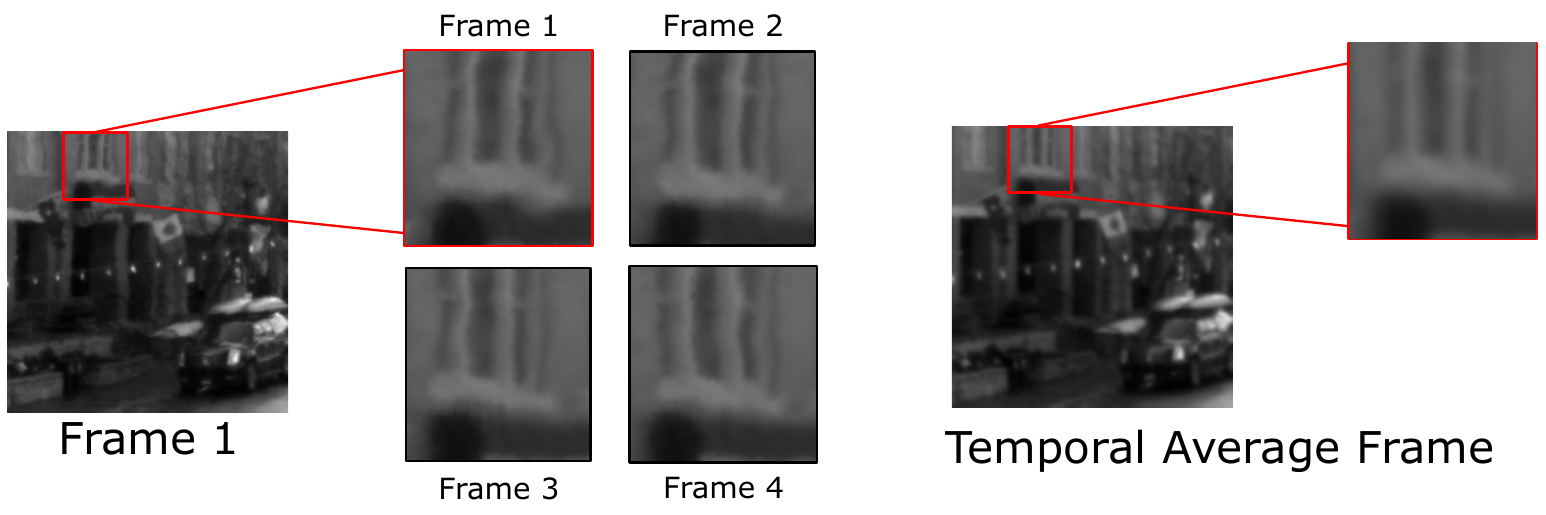}
\end{tabular}
\end{center}
\caption{ \label{fig:tilt_reduction}
Effect of temporal averaging in reducing frame tilt and geometric distortion across multiple captures of the same scene.}
\end{figure}

As shown in Fig.~\ref{fig:tilt_reduction}, temporal averaging helps reduce random tilt and geometric distortions across the sequence, making consistent image features more accessible to the network.

This idea was implemented after evaluating the baseline Transformer performance without the additional frame. Preliminary results suggest that including the averaged frame improves the network’s focus on stable structures, resulting in better restoration quality.

The updated input preparation process is summarized as follows:
\begin{itemize}
    \item Load $n$ consecutive frames from the dataset under identical turbulence conditions.
    \item Compute a pixel-wise average of the $n$ frames.
    \item Stack the original $n$ frames with the averaged frame, producing a $(n+1)$-frame input.
\end{itemize}

This approach enriches the spatial context available to the model and improves its ability to suppress frame-specific distortions, ultimately leading to more robust restoration performance.

The use of a temporal average is motivated by the observation that atmospheric turbulence introduces predominantly random frame-to-frame distortions, whereas the underlying scene remains unchanged. By averaging multiple observations of the same scene, random fluctuations tend to cancel while the persistent structural information is reinforced, producing a more stable approximation of the latent scene. This averaged frame is therefore not intended to reconstruct the ground-truth image directly, but rather to provide the Transformer with an additional structural prior that facilitates feature extraction and temporal consistency during restoration.

Alternative formulations could also be considered, including weighted temporal averaging, median-based aggregation, optical-flow-aligned averaging, or attention-based frame fusion. These approaches may further improve the quality of the structural prior and represent interesting directions for future work. In this study, the simple temporal average was selected because of its computational efficiency, ease of implementation, and its ability to provide consistent improvements without introducing additional trainable parameters.

\subsection{3D Mamba-Based Atmospheric Turbulence Removal (MAMAT)}
In addition to the GAN and Transformer based models trained in this study, we include MAMAT ~\cite{hill2025mamat} algorithm as an external reference method. The MAMAT model is a recently proposed learning-based approach that leverages structured state-space modeling through the 3D Mamba architecture to address spatiotemporal distortions in turbulence-degraded video sequences.

The MAMAT framework employs a dual-module design. The first module performs non-rigid spatiotemporal alignment using deformable 3D convolutions to mitigate frame-to-frame spatial displacements induced by atmospheric turbulence. The second module focuses on visual enhancement and detail preservation, exploiting the efficiency of 3D Mamba blocks to model long-range spatiotemporal dependencies. This combination enables MAMAT to jointly stabilize and enhance distorted video content.

In this work, MAMAT is retrained using the SOTIS dataset without any architectural modification and is used solely as a reference baseline for comparison. Its inclusion provides an additional point of reference against state-of-the-art sequence-based turbulence mitigation methods, complementing the comparison with Transformer-based models presented later in the Results section.

\section{Results}

\subsection{GAN Results}\label{sec:gan_results}

The performance of the GAN model was evaluated under varying atmospheric turbulence conditions using stabilized input images, as described in the Methodology section, ~\ref{sec:metodologia}. Prior to training and inference, all turbulence-degraded frames were preprocessed using stabilization techniques to reduce geometric distortions and enforce spatial alignment with the corresponding ground truth images. As a result, the GAN is primarily tasked with restoring residual blur and photometric degradation rather than learning geometric correction.

\begin{figure}
\begin{center}
\begin{tabular}{c}
\includegraphics[height=9.5cm]{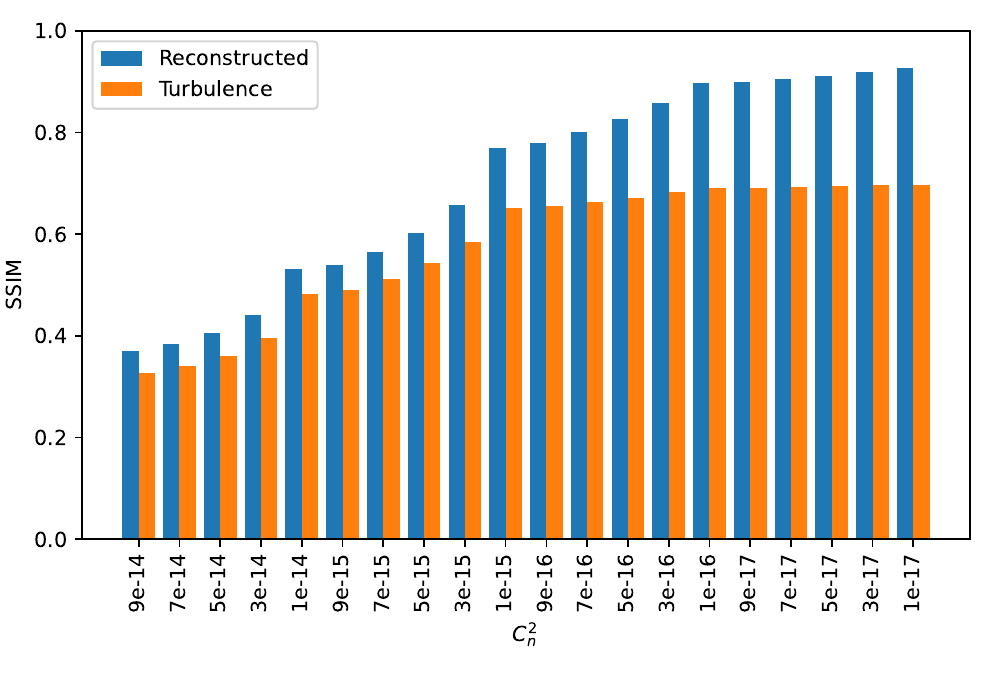}
\end{tabular}
\end{center}
\caption{ \label{fig:turbC}
Performance of the GAN model as a function of turbulence strength $C_n^2$.}
\end{figure}

Figure~\ref{fig:turbC} shows the reconstruction quality, measured by SSIM, as a function of turbulence strength $C_n^2$. Under weak turbulence conditions, the GAN achieves high reconstruction quality, with SSIM values approaching 0.9. As the turbulence strength increases, performance degrades progressively, with reconstructed SSIM values decreasing to approximately 0.7 for stronger turbulence regimes. This trend reflects the increasing difficulty of recovering high-frequency details as residual distortions become more severe, even after stabilization.

\begin{figure}
\begin{center}
\begin{tabular}{c}
\includegraphics[height=6.5cm]{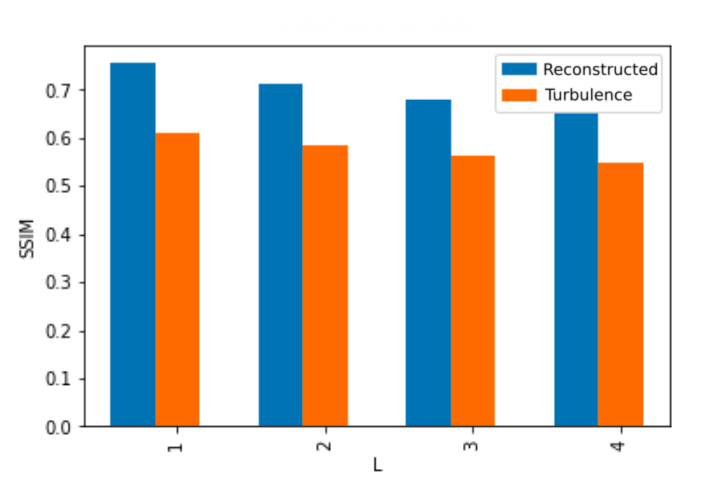}
\end{tabular}
\end{center}
\caption{ \label{fig:distance}
Performance of the GAN model with respect to sensor–target distance \( L \).}
\end{figure}

A similar trend is observed when the sensor-to-target distance \( L \) is varied. As shown in Fig.~\ref{fig:distance}, reconstruction quality improves when the sensor is closer to the target, and degrades with increasing distance. This again suggests the GAN’s limited ability to recover structural content under stronger geometric distortion.

It is important to emphasize that, although stabilization mitigates large-scale geometric distortions such as tilt, the GAN itself operates on single-frame inputs and does not explicitly model temporal dependencies or geometric motion. Consequently, the role of stabilization is critical in enabling effective GAN restoration. Without this preprocessing step, preliminary experiments showed unstable training and poor reconstruction quality. These observations highlight a fundamental limitation of single-frame adversarial models for turbulence mitigation: their reliance on external alignment mechanisms to compensate for geometric distortions.

To further analyze the behavior of the GAN model, the dataset was stratified according to the SSIM of the turbulence-degraded inputs. Table~\ref{tab:tabla1} reports the average SSIM before and after reconstruction across different SSIM bins. The results show that the reconstruction performance varies considerably with the initial image quality, with the largest absolute SSIM gain (0.1790) observed in the $(0.6\text{--}0.7]$ interval. This stratified analysis provides a more detailed characterization of the GAN performance across different turbulence regimes than the global averages presented previously.

\begin{table}[ht]
\caption{GAN model performance across SSIM bins. The best improvement occurs in the [0.6–0.7] range.}
\label{tab:tabla1}
\begin{center}       
\begin{tabular}{|l|c|c|c|}
\hline
\rule[-1ex]{0pt}{3.5ex} SSIM Bin & Turbulence (SSIM) & Reconstructed (SSIM) & SSIM Gain \\
\hline\hline
\rule[-1ex]{0pt}{3.5ex} (0–0.2]     & 0.1823 & 0.2284 & 0.0461  \\
\hline
\rule[-1ex]{0pt}{3.5ex} (0.2–0.3]   & 0.2561 & 0.3023 & 0.0462  \\
\hline
\rule[-1ex]{0pt}{3.5ex} (0.3–0.4]   & 0.3520 & 0.4058 & 0.0538  \\
\hline
\rule[-1ex]{0pt}{3.5ex} (0.4–0.5]   & 0.4506 & 0.5236 & 0.0730  \\
\hline
\rule[-1ex]{0pt}{3.5ex} (0.5–0.6]   & 0.5534 & 0.6863 & 0.1329  \\
\hline
\rule[-1ex]{0pt}{3.5ex} (0.6–0.7]   & 0.6544 & 0.8334 & \textbf{0.1790 } \\
\hline
\rule[-1ex]{0pt}{3.5ex} (0.7–0.8]   & 0.7427 & 0.8822 & 0.1395  \\
\hline
\rule[-1ex]{0pt}{3.5ex} (0.8–1.0]   & 0.8350 & 0.9091 & 0.0741  \\
\hline
\end{tabular}
\end{center}
\end{table}

The distribution of SSIM gains across the different bins reveals that the GAN exhibits its highest restoration capability under intermediate turbulence conditions rather than at the two extremes of the degradation spectrum. When the initial SSIM is very low ((0--0.2]), severe geometric distortions and residual blur remain even after stabilization, limiting the amount of recoverable information available to the generator. Conversely, when the input images already exhibit high structural similarity ((0.8--1.0]), only minor degradations remain, leaving relatively little room for measurable improvement. Consequently, the largest SSIM gains are achieved in the intermediate bins, where stabilization has successfully reduced the dominant geometric distortions while sufficient blur and photometric degradation remain for the generator to effectively restore.

\subsection{Results for the GEN-only model}
This subsection evaluates the generator-only (GEN) models introduced in Section ~\ref{sec:GEN-only}, which were trained without adversarial feedback and optimized using the combined MSE+SSIM supervised loss. As in the full GAN setting, GEN operates on stabilized single-frame inputs; therefore, the quality of the restoration depends not only on the network’s reconstruction capability, but also on the stabilization method used during preprocessing.

To quantify the impact of stabilization on GEN performance, we trained one GEN model per stabilization strategy (NLTV-LK, NLTV-TVL1, TV-LK, TV-TVL1, and Temporal Average), as well as an additional GEN model trained using all stabilization methods combined. For evaluation, the test set is stratified according to the SSIM of the turbulence-degraded inputs, enabling a consistent comparison across distortion severity. 

Table~\ref{tab:gen_bins_by_stab} summarizes the average SSIM within each input SSIM bin (``Turbulence'' column) and reports the reconstructed SSIM obtained by GEN under each stabilization method. This format allows direct comparison of reconstruction quality as a function of both (i) the initial distortion level and (ii) the stabilization strategy used to generate the GEN input. The best-performing stabilization method is then analyzed in more detail using the $C_n^2$ and $L$ performance curves, analogous to the GAN results reported in Section~\ref{sec:gan_results}.

\begin{table}[ht]
\caption{The performance of the GEN-only model across SSIM bins for different stabilization-based training configurations is shown. Each column corresponds to a GEN model trained using inputs generated by a specific stabilization method, as well as a mixed training configuration that uses all stabilization methods jointly.}

\label{tab:gen_bins_by_stab}
\centering
\small
\begin{tabular}{|l|c|c|c|c|c|c|c|}
\hline
SSIM B & Turb & NLTV--LK & TV--LK & TV--TVL1 & T.A. & NLTV--TVL1 & all \\
\hline\hline
(0--0.2]   & 0.1748 & 0.1606 & 0.1608 & 0.1616 & 0.3321 & 0.3326 & 0.2084 \\
\hline
(0.2--0.3] & 0.2577 & 0.2537 & 0.2540 & 0.2481 & 0.4142 & 0.4114 & 0.2989 \\
\hline
(0.3--0.4] & 0.3511 & 0.3895 & 0.3898 & 0.3585 & \textbf{0.5022} & \textbf{0.5021} & 0.4204 \\
\hline
(0.4--0.5] & 0.4506 & 0.5627 & 0.5630 & 0.5014 & 0.6306 & 0.6299 & 0.5795 \\
\hline
(0.5--0.6] & 0.5554 & 0.7611 & 0.7613 & 0.7079 & 0.7849 & 0.7848 & 0.7587 \\
\hline
(0.6--0.7] & 0.6525 & 0.8647 & 0.8648 & 0.8436 & 0.8694 & 0.8694 & 0.8570 \\
\hline
(0.7--0.8] & 0.7390 & 0.8920 & 0.8921 & 0.8824 & 0.8937 & 0.8937 & 0.8864 \\
\hline
(0.8--1.0] & 0.8316 & 0.9231 & 0.9231 & 0.9196 & 0.9239 & 0.9239 & 0.9188 \\
\hline
\end{tabular}
\end{table}
Table~\ref{tab:gen_bins_by_stab} highlights the strong dependence of the GEN-only model on the choice of stabilization strategy and, importantly, on the \emph{training regime} used to generate its inputs. Among the individual stabilization methods, temporal averaging (T.A.) and NLTV-TVL1 consistently achieve the highest reconstruction quality across all SSIM bins, with both methods producing substantial gains over the turbulence-degraded inputs, particularly in the low- and mid-SSIM regimes. For example, in the $(0.3-0.4]$ bin, the reconstructed SSIM increases from 0.3511 to approximately 0.50 when using either T.A.\ or NLTV-TVL1, compared to more modest improvements obtained with other stabilization methods.

The all stabilization configuration should be interpreted differently from the individual stabilization methods. Rather than representing a single preprocessing strategy, it corresponds to a multi stabilization training regime in which the GEN model is trained using inputs generated by all stabilization methods jointly. As shown in Table~\ref{tab:gen_bins_by_stab}, this configuration does not consistently achieve the highest SSIM values. However, the all configuration maintains reconstruction quality close to these best performing methods while avoiding sensitivity to a specific stabilization pipeline. This behavior indicates that training with multiple stabilized inputs improves robustness and generalization, even if it does not maximize SSIM for any single turbulence regime.

From a methodological perspective, these results indicate that GEN-only training benefits not only from effective stabilization, but also from diversity in stabilized inputs. Temporal averaging provides a strong structural prior by suppressing random frame-to-frame jitter, while NLTV-TVL1 combines nonlocal regularization with robust optical flow estimation, yielding improved alignment in textured and high-contrast regions. When all stabilization methods are used jointly during training, the generator is exposed to a broader distribution of residual artifacts and alignment errors, reducing sensitivity to any single preprocessing pipeline and resulting in the most robust overall performance.

\begin{figure}
\begin{center}
\begin{tabular}{c}
\includegraphics[height=9.2cm]{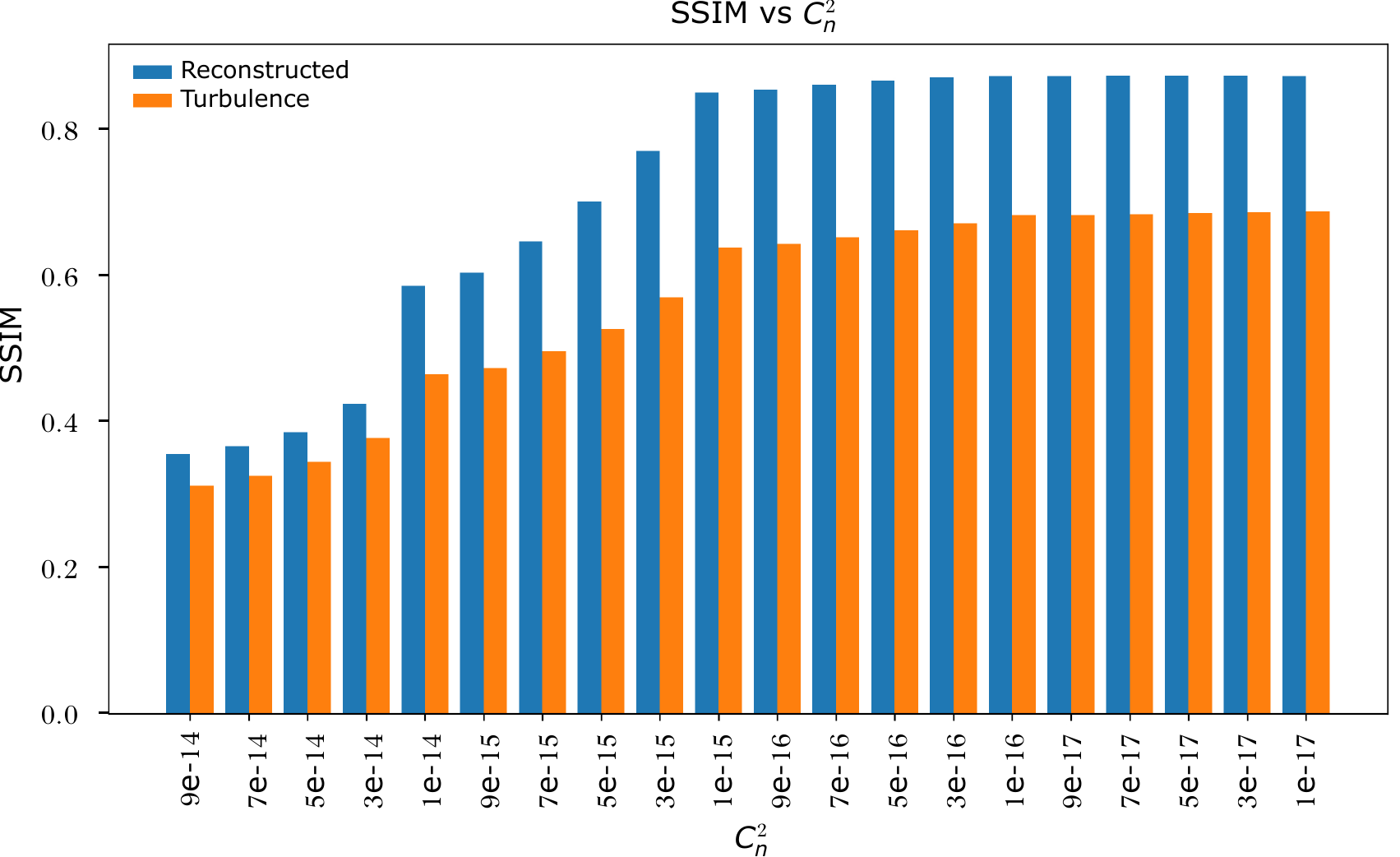}
\end{tabular}
\end{center}
\caption{ \label{fig:turbCGenAll}
Performance of the GEN-only model as a function of turbulence strength $C_n^2$ using all stabilization methods.}
\end{figure}

\begin{figure}
\begin{center}
\begin{tabular}{c}
\includegraphics[height=5.9cm]{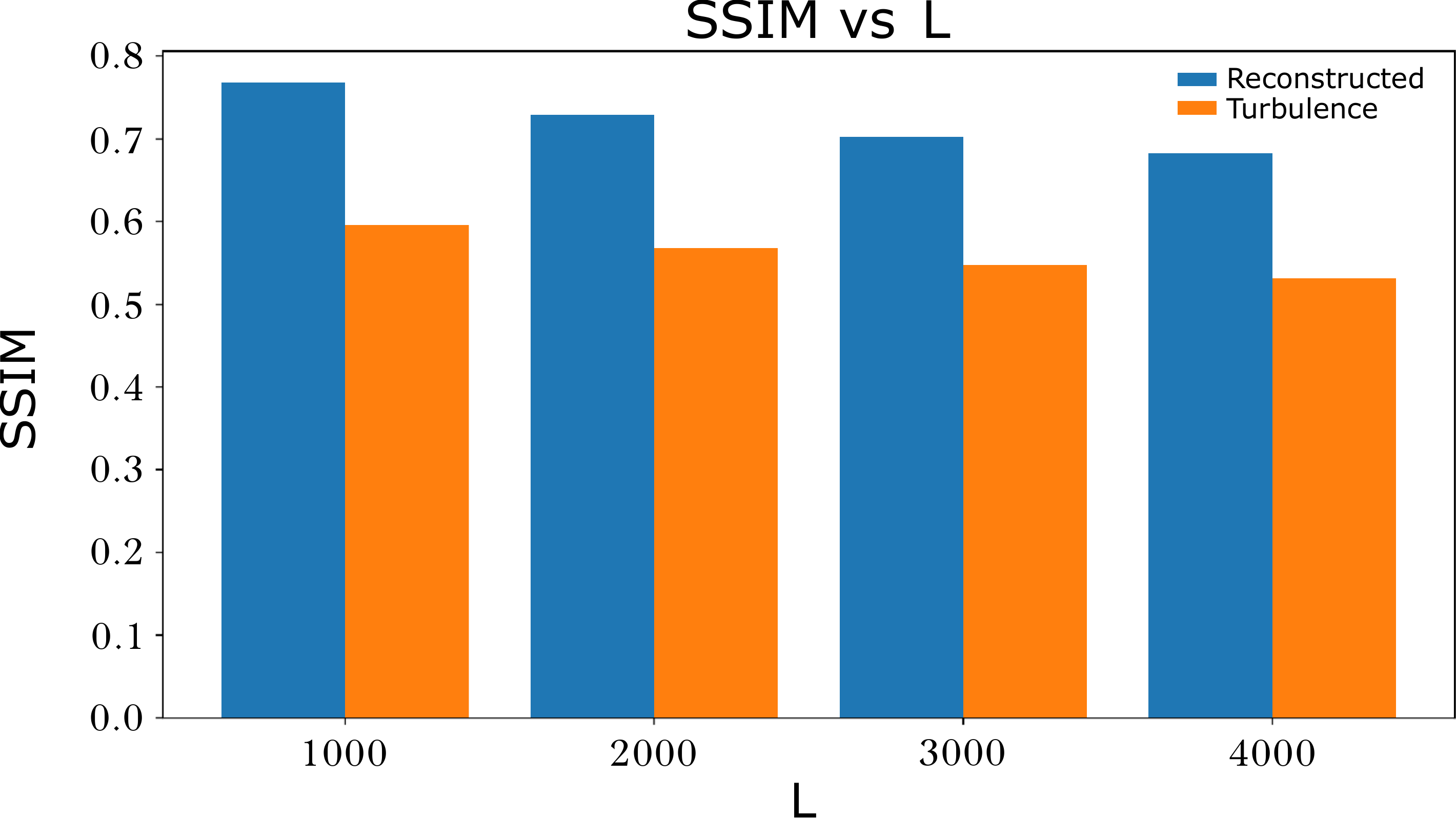} 
\end{tabular}
\end{center}
\caption{ \label{fig:distanceGenall}
Performance of the GEN-only model with respect to sensor–target distance \( L \) using all stabilization methods.}
\end{figure}

Figures~\ref{fig:turbCGenAll} and~\ref{fig:distanceGenall} report the performance of the GEN-only model trained using the all stabilization configuration, which was identified in Table~\ref{tab:gen_bins_by_stab}. Figure~\ref{fig:turbCGenAll} shows the reconstructed SSIM as a function of turbulence strength $C_n^2$. Across all turbulence levels, the GEN model trained with all stabilization methods consistently outperforms the turbulence-degraded inputs, with particularly pronounced gains in the low and mid turbulence regimes. As $C_n^2$ increases, reconstruction quality gradually degrades, reflecting the increasing severity of residual distortions; however, the performance gap between reconstructed and turbulent images remains substantial, indicating robust generalization.

A similar trend is observed when varying the sensor--target distance $L$, as shown in Fig.~\ref{fig:distanceGenall}. Reconstruction quality decreases as the distance increases, corresponding to stronger turbulence effects, yet the GEN model maintains a clear improvement over the input across all distances. The relatively smooth degradation with respect to both $C_n^2$ and $L$ suggests that training with all stabilization methods enables the generator to learn restoration features that are less sensitive to a specific distortion regime or preprocessing artifact.

The trends observed in Figs.~\ref{fig:turbCGenAll} and~\ref{fig:distanceGenall} are consistent with the behavior previously identified in the SSIM-bin analysis. Under weak and moderate turbulence, stabilization removes a large portion of the geometric distortions, allowing the generator to focus primarily on recovering residual blur and structural details. As the turbulence strength increases, however, the quality of the stabilized input gradually deteriorates, limiting the amount of recoverable information available to the network. Consequently, the reconstruction quality decreases smoothly with increasing $C_n^2$ and sensor--target distance, while consistently remaining above the turbulence-degraded baseline.

An important observation is that the performance degradation is gradual rather than abrupt. Unlike the full GAN configuration, which relies on adversarial supervision, the GEN-only model is optimized exclusively through reconstruction losses. The smooth evolution of the SSIM curves with both turbulence strength and propagation distance suggests that the learned mapping remains stable across different distortion levels. This behavior indicates that removing the discriminator improves optimization stability without substantially reducing the model's ability to generalize across the turbulence conditions represented in the training dataset.

\subsection{Transformer Results}
The same set of experiments was conducted using the Transformer-based model. In contrast to the GAN, the Transformer was trained and evaluated on images that include both blur and tilt distortions. This extended scope demonstrates the model’s capability to handle more complex and realistic atmospheric turbulence conditions, making the evaluation more representative of real-world scenarios.

We begin by evaluating the performance of the TMT architecture, as introduced in prior work, trained and evaluated on the SOTIS dataset without architectural modification. This experiment serves as a reference baseline within the Transformer-based family and provides a consistent point of comparison for the enhanced TMT variants introduced later in this section. By training the original TMT under the same dataset conditions and evaluation protocol used throughout this study, we ensure that any observed performance differences can be attributed to architectural and input-level modifications rather than dataset bias.

\begin{table}[ht]
\caption{Original transformer model performance across SSIM bins. The strongest improvements occur in the [0.5–0.6] and [0.6–0.7] ranges.}
\label{tab:tablaTransformerOriginal}
\begin{center}       
\begin{tabular}{|l|c|c|c|}
\hline
\rule[-1ex]{0pt}{3.5ex} SSIM Bin & Turbulence (SSIM) & Reconstructed (SSIM) & SSIM Gain \\
\hline\hline
\rule[-1ex]{0pt}{3.5ex} (0–0.2]     & 0.1654 & 0.2274 & 0.0618 \\
\hline
\rule[-1ex]{0pt}{3.5ex} (0.2–0.3]   & 0.2535 & 0.3470 & 0.0934 \\
\hline
\rule[-1ex]{0pt}{3.5ex} (0.3–0.4]   & 0.3518 & 0.4681 & \textbf{0.1164} \\
\hline
\rule[-1ex]{0pt}{3.5ex} (0.4–0.5]   & 0.4529 & 0.5707 & \textbf{0.1178} \\
\hline
\rule[-1ex]{0pt}{3.5ex} (0.5–0.6]   & 0.5533 & 0.6562& 0.1029 \\
\hline
\rule[-1ex]{0pt}{3.5ex} (0.6–0.7]   & 0.6465 & 0.7288 & 0.0824 \\
\hline
\rule[-1ex]{0pt}{3.5ex} (0.7–0.8]   & 0.7393 & 0.8050 & 0.0657 \\
\hline
\rule[-1ex]{0pt}{3.5ex} (0.8–1.0]   & 0.8426 & 0.8868 & 0.0443 \\
\hline
\end{tabular}
\end{center}
\end{table}

Table~\ref{tab:tablaTransformerOriginal} summarizes the SSIM performance of the original TMT architecture across different turbulence severity levels. The results show that, even without architectural or input-level modifications, TMT provides consistent reconstruction gains across all SSIM bins, confirming the effectiveness of multi-frame attention for turbulence mitigation. The largest improvements are observed in the low to mid SSIM regimes, particularly in the $(0.3-0.5]$ range, where the largest SSIM gain exceeds 0.11. This behavior indicates that the original TMT is especially effective when input images exhibit moderate levels of geometric and photometric distortion.

As the input SSIM increases, the absolute SSIM gain gradually decreases, with gains below 0.05 observed in the highest SSIM bin $(0.8-1.0]$. This trend suggests diminishing returns when the input images are already relatively clean, a behavior consistent with other restoration models. Nonetheless, the reconstructed SSIM remains substantially higher than the turbulence-degraded baseline across all bins, demonstrating robust performance even under challenging conditions.

An additional observation is that the SSIM gains vary smoothly across consecutive bins, without abrupt changes in performance. This gradual evolution indicates that the original TMT responds consistently as turbulence severity changes, rather than exhibiting abrupt performance degradation at specific distortion levels. Such behavior is desirable for practical applications, where atmospheric turbulence typically varies continuously rather than transitioning abruptly between discrete regimes.

These baseline results establish a strong reference point for evaluating the impact of the architectural and input-level enhancements introduced in the following experiments. In particular, they provide a quantitative benchmark against which improvements such as multi-scale refinements and the inclusion of a temporally averaged frame can be assessed in a controlled and interpretable manner.

\begin{figure}
\begin{center}
\begin{tabular}{c}
\includegraphics[height=10.0cm]{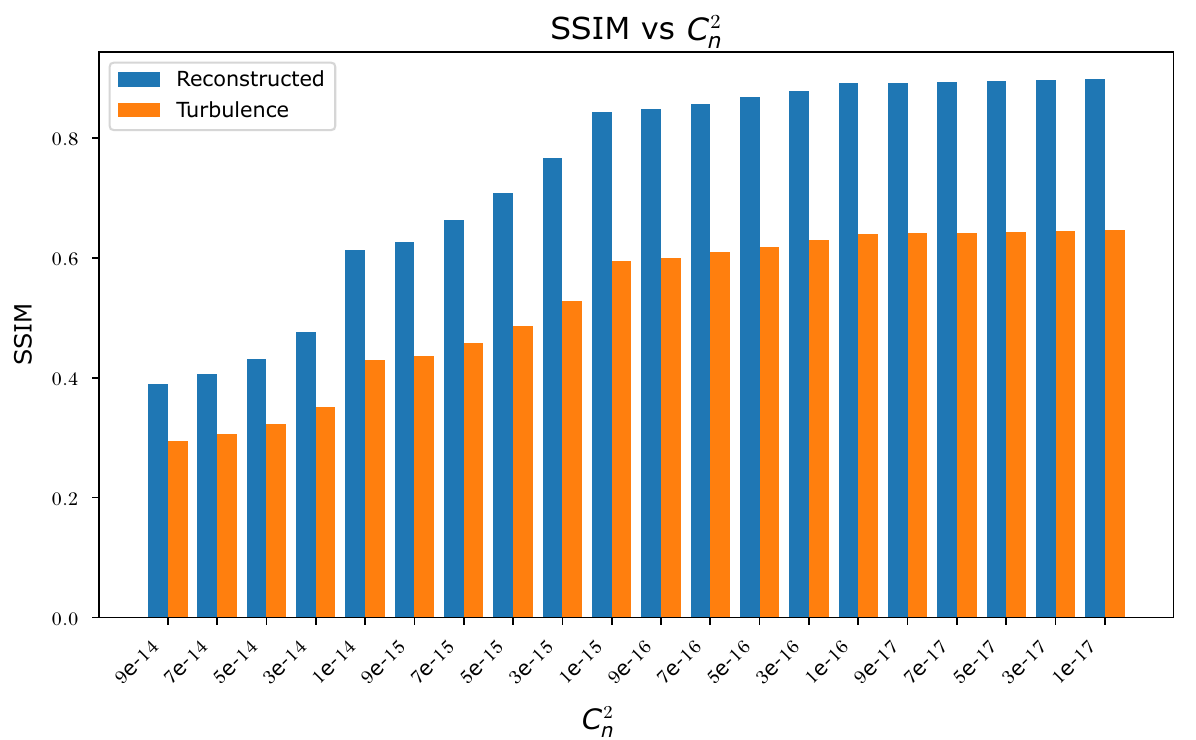}
\end{tabular}
\end{center}
\caption{ \label{fig:turbCTransformers}
Performance of the Transformer model with respect to turbulence strength $C_n^2$.}
\end{figure}

Having established the baseline performance of the original TMT architecture, we now evaluate the \textit{modified TMT model} introduced in this work. This version incorporates architectural refinements and training enhancements described in Section ~\ref{sec:metodologia}, while maintaining the same dataset and evaluation protocol. Figure~\ref{fig:turbCTransformers} shows the performance of this modified Transformer across different turbulence strengths, as measured by $C_n^2$. The model achieves higher SSIM scores under mild turbulence and maintains relatively stable performance as distortion increases. The performance drop observed at high $C_n^2$ values is expected due to the increased difficulty of restoring severely degraded images.

\begin{figure}
\begin{center}
\begin{tabular}{c}
\includegraphics[scale=0.55]{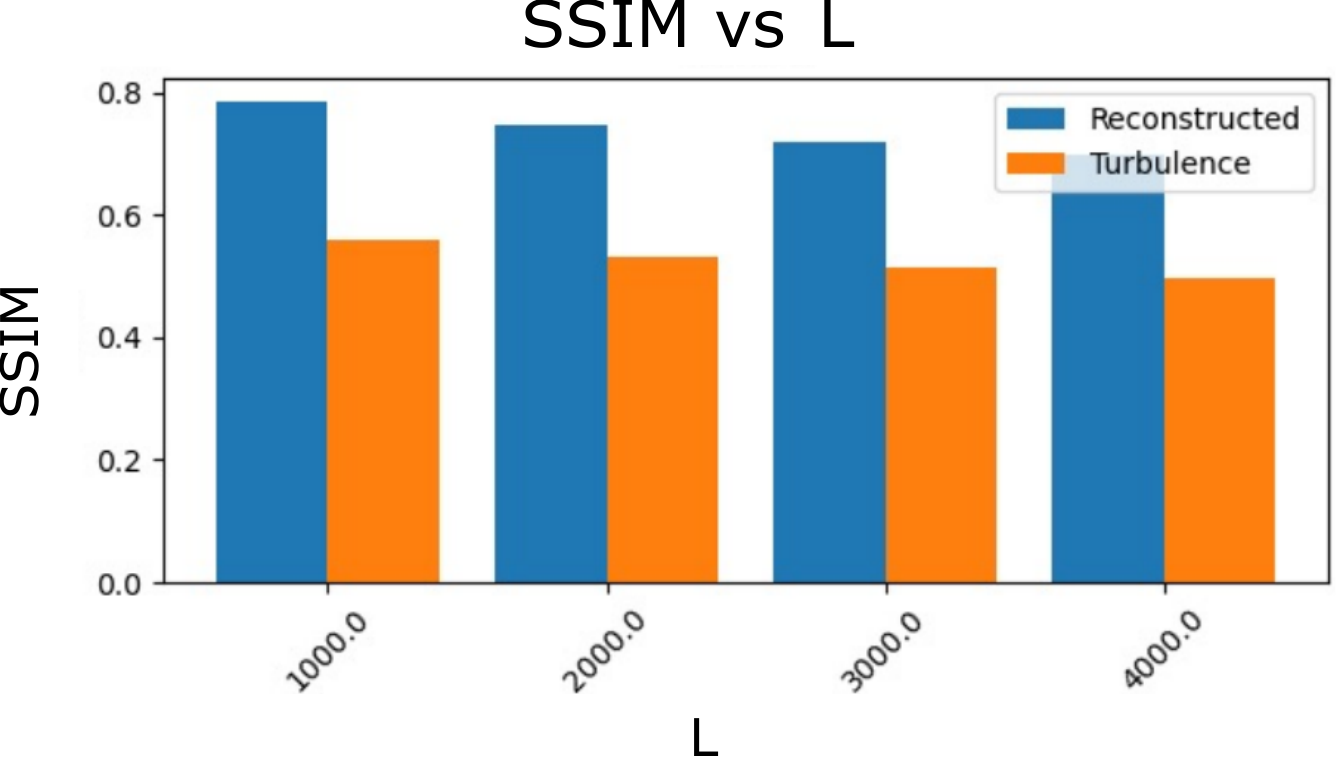}
\end{tabular}
\end{center}
\caption{ \label{fig:distanceTransformers}
Performance of the Transformer model as a function of sensor–target distance \( L \).}
\end{figure}

A similar trend is observed when varying the sensor-to-target distance \( L \). As illustrated in Fig.~\ref{fig:distanceTransformers}, reconstruction quality improves at shorter distances and degrades as distance increases, corresponding to stronger turbulence effects.

Table~\ref{tab:tabla2} summarizes the Transformer’s SSIM performance across different turbulence bins. The greatest improvements are observed in the [0.5–0.7] range, with SSIM gains exceeding 0.25. Even in the most challenging low-SSIM cases ([0–0.3]), the Transformer achieves meaningful reconstruction gains, demonstrating robustness across distortion levels.

\begin{table}[ht]
\caption{Modified TMT model performance across SSIM bins. The strongest improvements occur in the [0.5–0.6] and [0.6–0.7] ranges.}
\label{tab:tabla2}
\begin{center}       
\begin{tabular}{|l|c|c|c|}
\hline
\rule[-1ex]{0pt}{3.5ex} SSIM Bin & Turbulence (SSIM) & Reconstructed (SSIM) & SSIM Gain \\
\hline\hline
\rule[-1ex]{0pt}{3.5ex} (0–0.2]     & 0.1654 & 0.2472 & 0.0816 \\
\hline
\rule[-1ex]{0pt}{3.5ex} (0.2–0.3]   & 0.2535 & 0.3810 & 0.1274 \\
\hline
\rule[-1ex]{0pt}{3.5ex} (0.3–0.4]   & 0.3518 & 0.5323 & 0.1805 \\
\hline
\rule[-1ex]{0pt}{3.5ex} (0.4–0.5]   & 0.4529 & 0.7055 & 0.2526 \\
\hline
\rule[-1ex]{0pt}{3.5ex} (0.5–0.6]   & 0.5533 & 0.8423 & \textbf{0.2890} \\
\hline
\rule[-1ex]{0pt}{3.5ex} (0.6–0.7]   & 0.6465 & 0.9013 & 0.2548 \\
\hline
\rule[-1ex]{0pt}{3.5ex} (0.7–0.8]   & 0.7393 & 0.9208 & 0.1815 \\
\hline
\rule[-1ex]{0pt}{3.5ex} (0.8–1.0]   & 0.8426 & 0.9113 & 0.0688 \\
\hline
\end{tabular}
\end{center}
\end{table}

A visual analysis of the SSIM results leads to the following key observations:

\begin{enumerate}
    \item \textbf{Higher Reconstruction Quality:} Reconstructed images (blue bars) consistently outperform turbulence-distorted inputs (orange bars), confirming the model's effectiveness in reducing distortion.
    
    \item \textbf{Trend with Increasing Turbulence:} As \( C_n^2 \) increases, SSIM values decline for both input and output images, reflecting the greater difficulty of restoring heavily degraded scenes.
    
    \item \textbf{Diminishing Returns:} At low turbulence levels (e.g., \( 10^{-17} \leq C_n^2 \leq 10^{-15} \)), reconstructed SSIM remains near 0.9. However, at higher turbulence values (e.g., \( \geq 3 \times 10^{-15} \)), restoration performance begins to decline.
    
    \item \textbf{Robustness:} Even under severe turbulence (e.g., \( C_n^2 = 10^{-14} \)), the Transformer maintains significantly higher SSIM than the distorted input, indicating strong model robustness.
    
    \item \textbf{Performance Gap:} The difference between reconstructed and turbulent images is largest at moderate turbulence levels, suggesting that the model is most effective when distortions are strong but not extreme.
\end{enumerate}

Compared with the original TMT, the modified model improves reconstruction quality across every SSIM bin, indicating that the proposed training modifications provide consistent benefits rather than improving only a particular turbulence regime. The largest gains are observed in the intermediate SSIM bins, where sufficient structural information remains available while geometric distortions are still significant. This suggests that the proposed training strategy enhances the model's ability to exploit multi-frame information under conditions where both temporal consistency and structural recovery are important.

\subsection{Effect of Including a temporally Averaged Frame}

To evaluate the impact of incorporating a temporally averaged frame, an additional set of experiments was conducted using the same model architecture and dataset configuration described in previous sections. In this setup, the network input consists of 12 turbulence distorted frames along with a 13th frame, computed as the temporal average of the original sequence.

The goal was to determine whether this averaged frame, which provides a stable approximation of the underlying scene structure, could improve restoration quality. Table~\ref{tab:average_frame_results} presents the SSIM results across different turbulence levels, stratified by SSIM bins. The inclusion of the average frame consistently improves reconstruction performance across all bins, with the most significant gains observed in the [0.5–0.6] and [0.6–0.7] ranges.

\begin{table}[ht]
\caption{Performance of the modified TMT model with the inclusion of a temporally averaged frame (TMT-TA) across SSIM bins. The strongest improvements are observed in the mid-range SSIM bins.}
\label{tab:average_frame_results}
\begin{center}       
\begin{tabular}{|l|c|c|c|}
\hline
\rule[-1ex]{0pt}{3.5ex} SSIM Bin & Turbulence (SSIM) & Reconstructed (SSIM) & SSIM Gain \\
\hline\hline
\rule[-1ex]{0pt}{3.5ex} (0–0.2]     & 0.1659 & 0.2485 & 0.0826 \\
\hline
\rule[-1ex]{0pt}{3.5ex} (0.2–0.3]   & 0.2539 & 0.3945 & 0.1405 \\
\hline
\rule[-1ex]{0pt}{3.5ex} (0.3–0.4]   & 0.3518 & 0.5504 & 0.1986 \\
\hline
\rule[-1ex]{0pt}{3.5ex} (0.4–0.5]   & 0.4529 & 0.7275 & 0.2746 \\
\hline
\rule[-1ex]{0pt}{3.5ex} (0.5–0.6]   & 0.5532 & 0.8670 & \textbf{0.3138} \\
\hline
\rule[-1ex]{0pt}{3.5ex} (0.6–0.7]   & 0.6464 & 0.9210 & 0.2746 \\
\hline
\rule[-1ex]{0pt}{3.5ex} (0.7–0.8]   & 0.7394 & 0.9439 & 0.2045 \\
\hline
\rule[-1ex]{0pt}{3.5ex} (0.8–1.0]   & 0.8426 & 0.9711 & 0.1285 \\
\hline
\end{tabular}
\end{center}
\end{table}

An important observation is that the benefit provided by the temporally averaged frame is consistent across all evaluated SSIM bins, although its magnitude depends on the initial image quality. The largest SSIM gains are observed in the intermediate SSIM range, particularly between $(0.4\text{--}0.7]$, where the averaged frame provides a reliable estimate of the underlying scene while the individual frames still contain complementary structural information. Under these conditions, the additional reference frame helps the Transformer better exploit temporal redundancy, resulting in noticeably larger reconstruction gains than those obtained by the modified TMT alone.

The improvement becomes smaller toward both extremes of the SSIM spectrum for different reasons. In the lowest SSIM bin, severe turbulence limits the quality of both the individual observations and their temporal average, reducing the amount of useful structural information available to the network. Conversely, in the highest SSIM bin the distorted inputs are already close to the ground truth, leaving relatively little room for additional improvement. These results indicate that the temporally averaged frame is most beneficial when sufficient scene structure can be recovered from multiple observations while meaningful restoration is still required.

\begin{figure}
\begin{center}
\begin{tabular}{c}
\includegraphics[width=16cm]{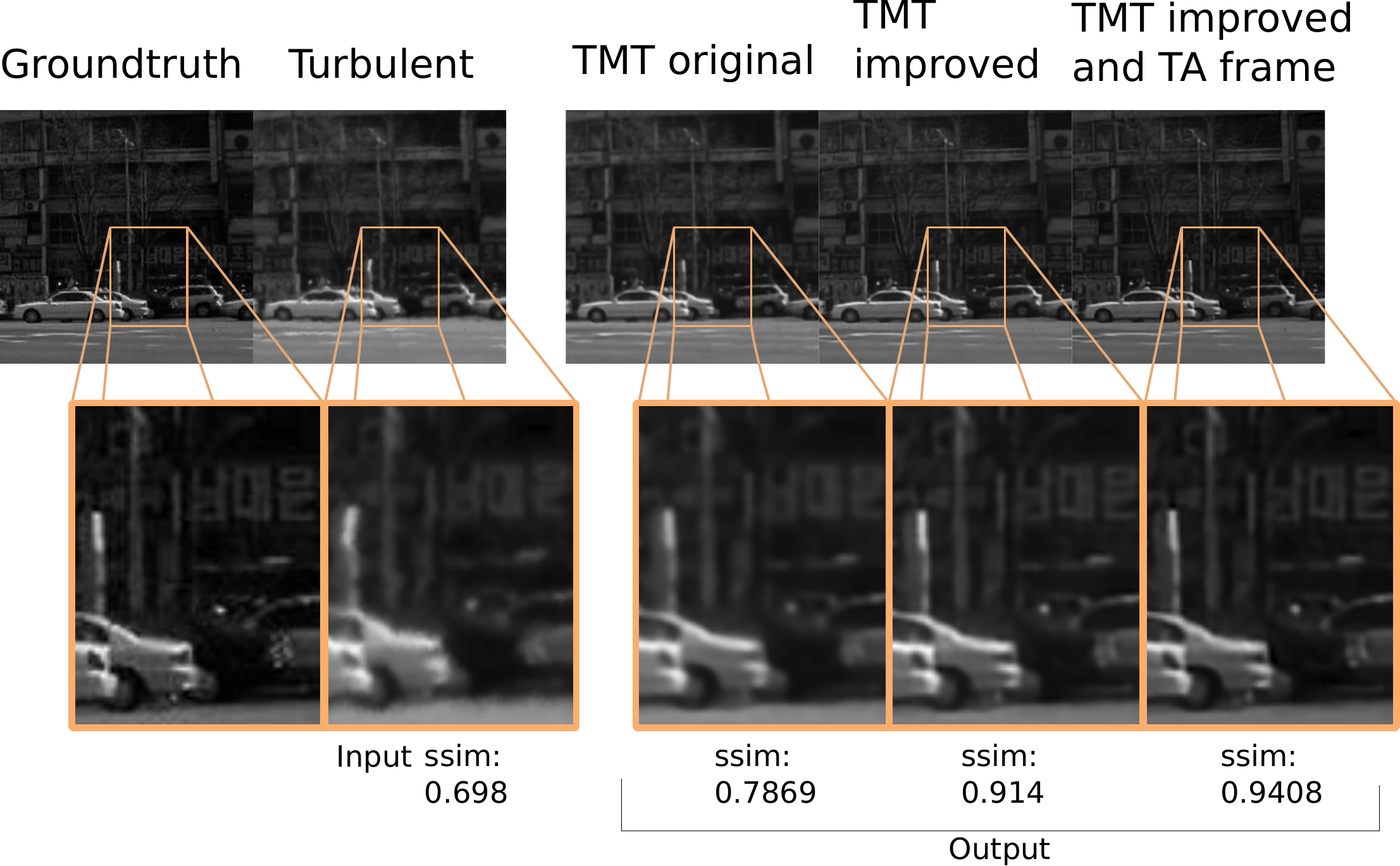}
\end{tabular}
\end{center}
\caption{ \label{fig:finalResultTMTMean}
Comparison of restoration results for the original TMT, the improved TMT, and the improved TMT with temporal average frame.}
\end{figure}
Figure~\ref{fig:finalResultTMTMean} presents a qualitative comparison between the improved TMT model and the improved TMT with temporal average frame. Both configurations significantly enhance structural clarity when compared to the turbulence degraded input. However, the version incorporating the temporally averaged frame exhibits visibly sharper edge definition and stronger contrast in high frequency regions, particularly around object boundaries and vertical structural elements.

The zoomed regions further highlight that the inclusion of the averaged frame improves local consistency and reduces residual blur, resulting in contours that more closely resemble the ground truth image. While the improved TMT already achieves substantial restoration quality, the additional temporal reference provides a subtle but consistent enhancement in edge sharpness and contrast, confirming the complementary role of temporal averaging within the Transformer based restoration framework.

\subsubsection{Evaluation on an External Dataset}

To further evaluate the robustness and generalization capability of the proposed Transformer model, additional experiments were conducted using the dataset employed by Zhang et al.~\cite{zhang2024imaging} for the evaluation of the original Turbulence Mitigation Transformer (TMT). Unlike the SOTIS dataset used throughout the primary experiments in this work, this dataset contains different image content and a broader distribution of atmospheric turbulence conditions, providing an independent benchmark for assessing cross-dataset performance.

The SOTIS dataset contains turbulence strengths defined by $C_n^2=a\times10^{-b}$ with $a\in\{1,3,5,7\}$ and $b\in\{14,15,16,17\}$. In contrast, the external dataset additionally includes intermediate turbulence coefficients ($a=2,4,6,8,$ and $9$) together with previously unseen turbulence magnitudes corresponding to exponent values of $10^{-13}$, $10^{-12}$, and $10^{-11}$. Consequently, the external dataset evaluates the proposed model under both interpolation conditions, where new coefficients are introduced within the turbulence regimes observed during training, and extrapolation conditions involving turbulence magnitudes outside the SOTIS training distribution.

Two evaluation protocols were considered. In the first experiment, the Transformer model trained exclusively on the SOTIS dataset was directly evaluated on the external dataset without any additional optimization. In the second experiment, transfer learning was performed by initializing the network with the weights learned from SOTIS and fine-tuning the model on the external dataset before evaluation.

\begin{figure}
\begin{center}
\begin{tabular}{c}
\includegraphics[width=16cm]{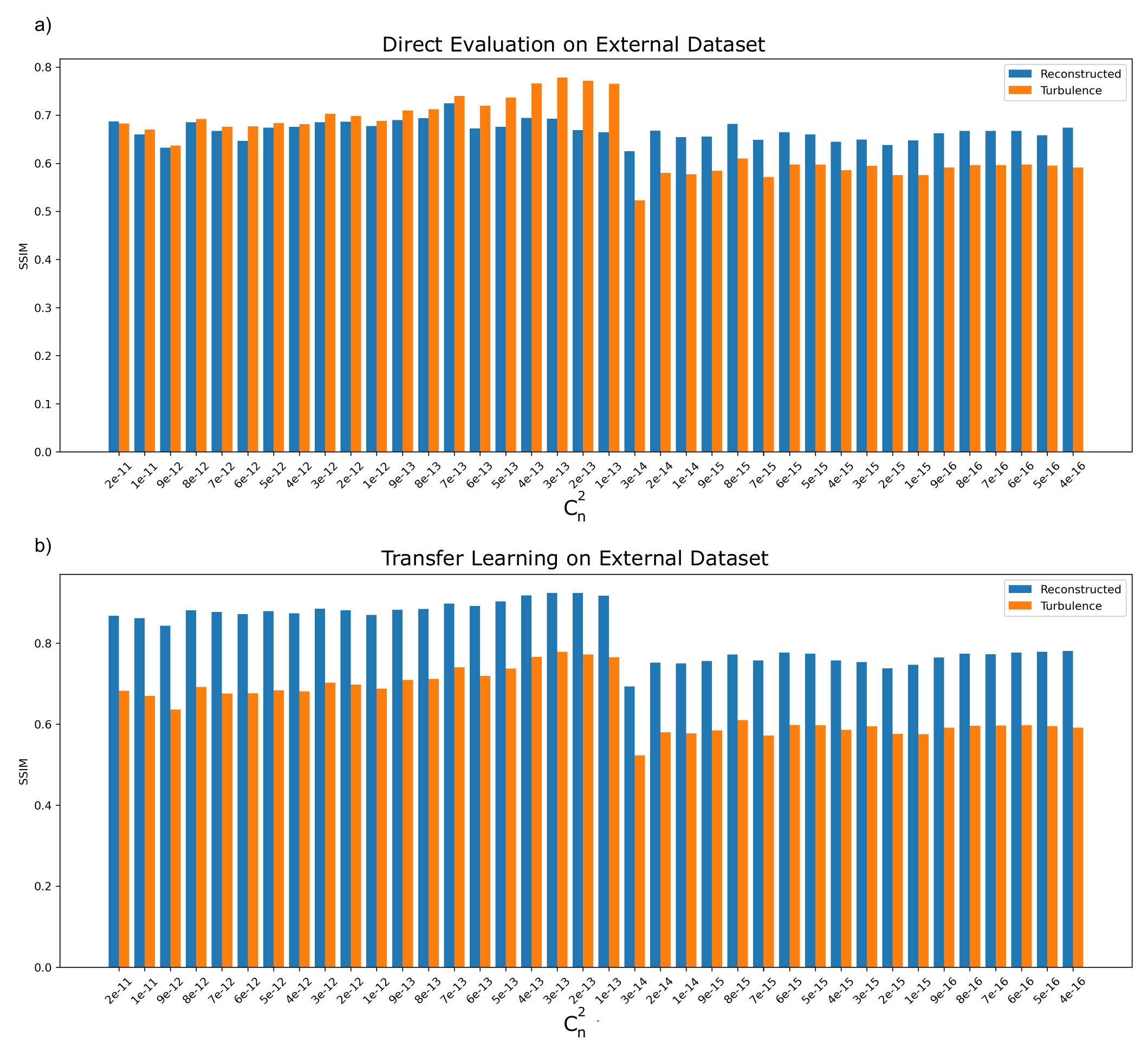}
\end{tabular}
\end{center}
\caption{ \label{fig:externalDatasetComparison}
Generalization performance of the proposed Transformer model on an external atmospheric turbulence dataset. (a) Direct evaluation using the model trained exclusively on the SOTIS dataset. (b) Performance after transfer learning on the external dataset.
}
\end{figure}

Figure~\ref{fig:externalDatasetComparison}(a) presents the direct evaluation of the proposed Transformer model on the external dataset without additional training. Although a performance reduction is observed due to the domain shift between datasets, the model maintains positive SSIM gains across a substantial portion of the evaluated turbulence conditions. A closer inspection reveals that reconstruction performance is considerably more sensitive to the order of magnitude of $C_n^2$ than to its multiplicative coefficient. When the exponent of $C_n^2$ matches those represented in the SOTIS dataset ($10^{-14}$, $10^{-15}$, and $10^{-16}$), the model consistently achieves positive reconstruction gains even for intermediate coefficients ($a=2,4,6,8,$ and $9$) that were never observed during training. In contrast, the largest degradation occurs for turbulence conditions associated with previously unseen exponents ($10^{-13}$, $10^{-12}$, and $10^{-11}$), indicating that the model generalizes well within the turbulence regimes represented during training but becomes less effective when evaluated on substantially different turbulence magnitudes.

Figure~\ref{fig:externalDatasetComparison}(b) shows the performance obtained after applying transfer learning on the external dataset. Compared with direct evaluation, reconstruction quality improves substantially across the entire range of turbulence conditions. The largest gains are observed for moderate and severe turbulence, while consistently high reconstruction quality is maintained under weaker turbulence conditions. These results demonstrate that the representations learned from the SOTIS dataset provide an effective initialization for adaptation to a different dataset, requiring only limited fine-tuning to recover strong restoration performance.

Overall, these experiments demonstrate that the proposed Transformer model exhibits a reasonable degree of cross-dataset generalization and can be efficiently adapted to new turbulence conditions through transfer learning, supporting the robustness of the proposed approach beyond the SOTIS dataset.

\begin{table}[ht]
\caption{Performance of the proposed Transformer model on the external dataset.}
\label{tab:external_dataset_results}
\begin{center}
\begin{tabular}{|l|c|c|c|c|}
\hline
\rule[-1ex]{0pt}{3.5ex}
SSIM Bin &
Turbulence &
Direct Evaluation &
Transfer Learning &
SSIM Gain \\
\hline\hline

(0--0.2] & 0.1582 & 0.2612 & 0.3467 & 0.1030 / 0.1885 \\
\hline
(0.2--0.3] & 0.2577 & 0.3773 & 0.4886 & 0.1196 / 0.2309 \\
\hline
(0.3--0.4] & 0.3539 & 0.4665 & 0.5927 & 0.1125 / \textbf{0.2388} \\
\hline
(0.4--0.5] & 0.4514 & 0.5435 & 0.6791 & 0.0922 / 0.2277 \\
\hline
(0.5--0.6] & 0.5515 & 0.6197 & 0.7609 & 0.0682 / 0.2094 \\
\hline
(0.6--0.7] & 0.6508 & 0.6878 & 0.8269 & 0.0370 / 0.1761 \\
\hline
(0.7--0.8] & 0.7472 & 0.7585 & 0.8838 & 0.0113 / 0.1366 \\
\hline
(0.8--1.0] & 0.8829 & 0.8638 & 0.9472 & -0.0192 / 0.0642 \\
\hline

\end{tabular}
\end{center}
\end{table}

Table~\ref{tab:external_dataset_results} provides a quantitative comparison between direct evaluation and transfer learning across the different SSIM bins. Without additional training, the proposed Transformer achieves positive SSIM gains in seven of the eight turbulence regimes, indicating that the learned representations retain a reasonable degree of generalization despite the domain shift between datasets. The only negative result is observed in the highest SSIM bin, where the difference between the two datasets has a greater impact on the refinement of already high-quality images.

After transfer learning, reconstruction quality improves consistently across all SSIM bins. Compared with direct evaluation, the SSIM gains increase substantially for every turbulence regime, with the largest improvements observed in the low- and intermediate-quality bins. These results indicate that the features learned from the SOTIS dataset provide a strong initialization, allowing the model to adapt efficiently to the external dataset with only limited fine-tuning.

Taken together, the results of Fig.~\ref{fig:externalDatasetComparison} and Table~\ref{tab:external_dataset_results} demonstrate that the proposed Transformer is capable of generalizing beyond the SOTIS dataset while remaining readily adaptable to previously unseen atmospheric turbulence conditions through transfer learning.

\subsection{MAMAT Results}

Table~\ref{tab:tablaResultadosMamat} summarizes the performance of the MAMAT algorithm across different SSIM bins when trained and evaluated on the SOTIS dataset. The results show that MAMAT provides consistent reconstruction improvements across all turbulence levels, confirming its ability to mitigate spatiotemporal distortions in turbulence degraded imagery. The strongest gains are observed in the low to mid SSIM regimes, particularly in the $(0.3-0.5]$ range, where the model achieves an absolute SSIM gain of approximately 0.06. This behavior indicates that MAMAT is most effective when the input images exhibit moderate levels of degradation.

As the input SSIM increases, the relative improvement gradually decreases, with gains dropping below four percent in the highest SSIM bins. This trend suggests diminishing returns when the input images are already relatively clean, a pattern that is consistent with other restoration approaches evaluated in this study. Nonetheless, the reconstructed SSIM remains consistently higher than the turbulence degraded baseline across all bins, demonstrating stable performance under a wide range of turbulence conditions.

When compared to the learning based models introduced in this work, the magnitude of the SSIM improvements achieved by MAMAT is more modest. This difference can be attributed to the fact that MAMAT is used as an external reference method without architectural modification or task specific adaptation to the SOTIS dataset. Despite this, its inclusion provides a valuable baseline for assessing the relative benefits of the proposed Transformer based enhancements, particularly in terms of structural fidelity and robustness under moderate turbulence.

\begin{table}[ht]
\caption{MAMAT performance across SSIM bins on the SOTIS dataset.}
\label{tab:tablaResultadosMamat}
\begin{center}       
\begin{tabular}{|l|c|c|c|}
\hline
\rule[-1ex]{0pt}{3.5ex} SSIM Bin & Turbulence (SSIM) & Reconstructed (SSIM) & SSIM Gain \\
\hline\hline
\rule[-1ex]{0pt}{3.5ex} (0–0.2]     & 0.1654 & 0.2255 & 0.0433 \\
\hline
\rule[-1ex]{0pt}{3.5ex} (0.2–0.3]   & 0.2535 & 0.3133 & 0.0523 \\
\hline
\rule[-1ex]{0pt}{3.5ex} (0.3–0.4]   & 0.3518 & 0.4128 & \textbf{0.0606} \\
\hline
\rule[-1ex]{0pt}{3.5ex} (0.4–0.5]   & 0.4529 & 0.5165 & \textbf{0.0647} \\
\hline
\rule[-1ex]{0pt}{3.5ex} (0.5–0.6]   & 0.5533 & 0.6116 & 0.0565 \\
\hline
\rule[-1ex]{0pt}{3.5ex} (0.6–0.7]   & 0.6465 & 0.6945 & 0.0463 \\
\hline
\rule[-1ex]{0pt}{3.5ex} (0.7–0.8]   & 0.7393 & 0.7710 & 0.0358 \\
\hline
\rule[-1ex]{0pt}{3.5ex} (0.8–1.0]   & 0.8426 & 0.8659 & 0.0316 \\
\hline
\end{tabular}
\end{center}
\end{table}

\subsubsection{Computational Complexity Analysis}

In addition to restoration quality, computational efficiency is an important consideration for the practical deployment of atmospheric turbulence mitigation algorithms. To provide a quantitative comparison, we evaluated the computational complexity of the proposed GAN, Generator-only (GEN), and Transformer (TMT) models in terms of architectural complexity and inference performance. All inference benchmarks were performed on a workstation equipped with an NVIDIA RTX A5000 GPU using a batch size of 1.

\begin{table}[ht]
\caption{Architectural complexity of the evaluated restoration models.}
\label{tab:complexity_training}
\begin{center}
\begin{tabular}{|l|r|r|}
\hline
\rule[-1ex]{0pt}{3.5ex}
Model &
Trainable Parameters &
Model Size (MB) \\
\hline\hline

GAN (training) &
812,434 &
3.10\\
\hline

GEN &
34,929 &
0.13\\
\hline

Transformer (TMT) &
22,936,370 &
87.50\\
\hline

\end{tabular}
\end{center}
\end{table}
\begin{table}[ht]
\caption{Inference performance of the evaluated restoration models.}
\label{tab:complexity_inference}
\begin{center}
\begin{tabular}{|l|c|r|r|r|}
\hline
\rule[-1ex]{0pt}{3.5ex}
Model &
Input &
Inference (ms/image) &
Throughput (img/s) \\
\hline\hline

GAN (training) &
$1\times256\times256$ &
2.10 &
475.17 \\
\hline

GEN &
$1\times256\times256$ &
2.04 &
489.90 \\
\hline

TMT &
$13\times120\times120$ &
753.60 &
1.33 \\
\hline

\end{tabular}
\end{center}
\end{table}

Tables~\ref{tab:complexity_training} and~\ref{tab:complexity_inference} summarize the architectural complexity and inference performance of the evaluated models, respectively. During training, the complete GAN requires both the generator and discriminator, resulting in a total of 812,434 trainable parameters. However, during deployment only the generator is used for image restoration, reducing the inference model to 34,929 trainable parameters with a model size of only 0.13~MB. The GEN-only configuration employs the same generator architecture and therefore exhibits nearly identical computational characteristics during inference.

In contrast, the Transformer model contains over 22.9 million trainable parameters and occupies approximately 87.5~MB. Furthermore, unlike the GAN-based approaches that process a single image, the Transformer simultaneously processes a sequence of 13 input frames using a multi-frame attention mechanism. Consequently, the Transformer requires substantially longer inference times (753.6~ms per image) and achieves a throughput of only 1.33 images per second, compared with approximately 2~ms per image and nearly 490 images per second for the generator-based models.

These results highlight the trade-off between computational efficiency and restoration performance. While the Transformer consistently achieves the highest restoration quality across most turbulence regimes, this improvement comes at the expense of significantly greater computational complexity and inference time. Conversely, the GAN and GEN-only configurations provide substantially faster inference and lower computational requirements, making them attractive alternatives for resource-constrained or real-time applications.

\subsection{Comparison}

This section presents a comparative analysis of the restoration performance of all evaluated algorithms across different turbulence regimes. The comparison considers both quantitative SSIM improvements and qualitative visual results, highlighting how different models perform under varying levels of image degradation.

\subsubsection{Quantitative Comparison (SSIM-based Evaluation)}

Tables~\ref{tab:tabla1} and \ref{tab:average_frame_results} further support this comparison when viewed through the lens of SSIM bin stratification. In this case, the GAN based results are contrasted with the Transformer model incorporating the temporally averaged frame. The TMT configuration with temporal averaging shows both more consistent and more significant SSIM improvements across most of the SSIM range. In particular, in the lowest SSIM bins $(0\text{--}0.2]$ and $(0.2\text{--}0.3]$, the Transformer with temporal averaging achieves an absolute SSIM gain of up to 0.1405, while the GAN model exhibits more modest gains of approximately 0.0462.

As the input SSIM increases, indicating cleaner inputs, both models exhibit smaller absolute SSIM gains. However, the Transformer continues to outperform the GAN across all SSIM bins. In the mid-range bins (0.4–0.5 and 0.5–0.6), the Transformer achieves SSIM gains ranging from 0.2746 to 0.3138, compared to 0.0730–0.1790 for the GAN. Even in the highest SSIM bin (0.8–1.0], where the input images are already relatively clean, the Transformer maintains an advantage, achieving an SSIM gain of 0.1285 compared to 0.0741 for the GAN, although the gains of both models decrease as the input quality improves.

These comparisons suggest that while the GAN model is effective at removing blur under mild turbulence conditions, the Transformer’s architecture and multi-frame attention mechanism make it substantially more capable in handling both geometric and photometric distortions. The Transformer emerges as a more versatile and robust solution for restoring atmospheric turbulence–affected imagery.

In addition to the comparison between GAN and Transformer based models, it is also instructive to directly contrast the best performing GEN configuration with the Transformer model incorporating a temporally averaged frame. Based on Table~\ref{tab:gen_bins_by_stab}, the GEN model trained with NLTV--TVL1 stabilization achieves the strongest performance among generator-only variants and serves as a representative upper bound for GEN based restoration.

When compared against this best GEN configuration, the relative performance of the TMT model with temporal averaging depends on the input quality regime. In the most challenging low SSIM range $(0\text{--}0.3]$, GEN with NLTV--TVL1 achieves reconstructed SSIM values between approximately 0.33 and 0.41, while the Transformer with temporal averaging reaches values between 0.25 and 0.39. These results indicate that under severe turbulence conditions, GEN-based restoration can be competitive and, in some cases, slightly outperform the Transformer approach.

However, as input quality improves, the performance gap becomes more pronounced in favor of the Transformer model. In the mid-range bins $(0.3\text{--}0.6]$, the Transformer achieves reconstructed SSIM values exceeding 0.72 and reaching up to 0.87, clearly surpassing the GEN configuration, which remains below these levels.

At higher SSIM bins $(0.6\text{--}1.0]$, the advantage of the Transformer with temporal averaging becomes even more evident, with reconstructed SSIM values consistently above 0.92, compared to GEN values that plateau around 0.89 to 0.92. These results indicate that while GEN-based restoration with strong stabilization can provide competitive performance in extreme turbulence conditions, the Transformer model with temporal averaging offers superior and more consistent reconstruction quality across moderate and mild turbulence regimes, reinforcing its robustness as a general-purpose solution.

\begin{figure}
\begin{center}
\begin{tabular}{c}
\includegraphics[height=5.5cm]{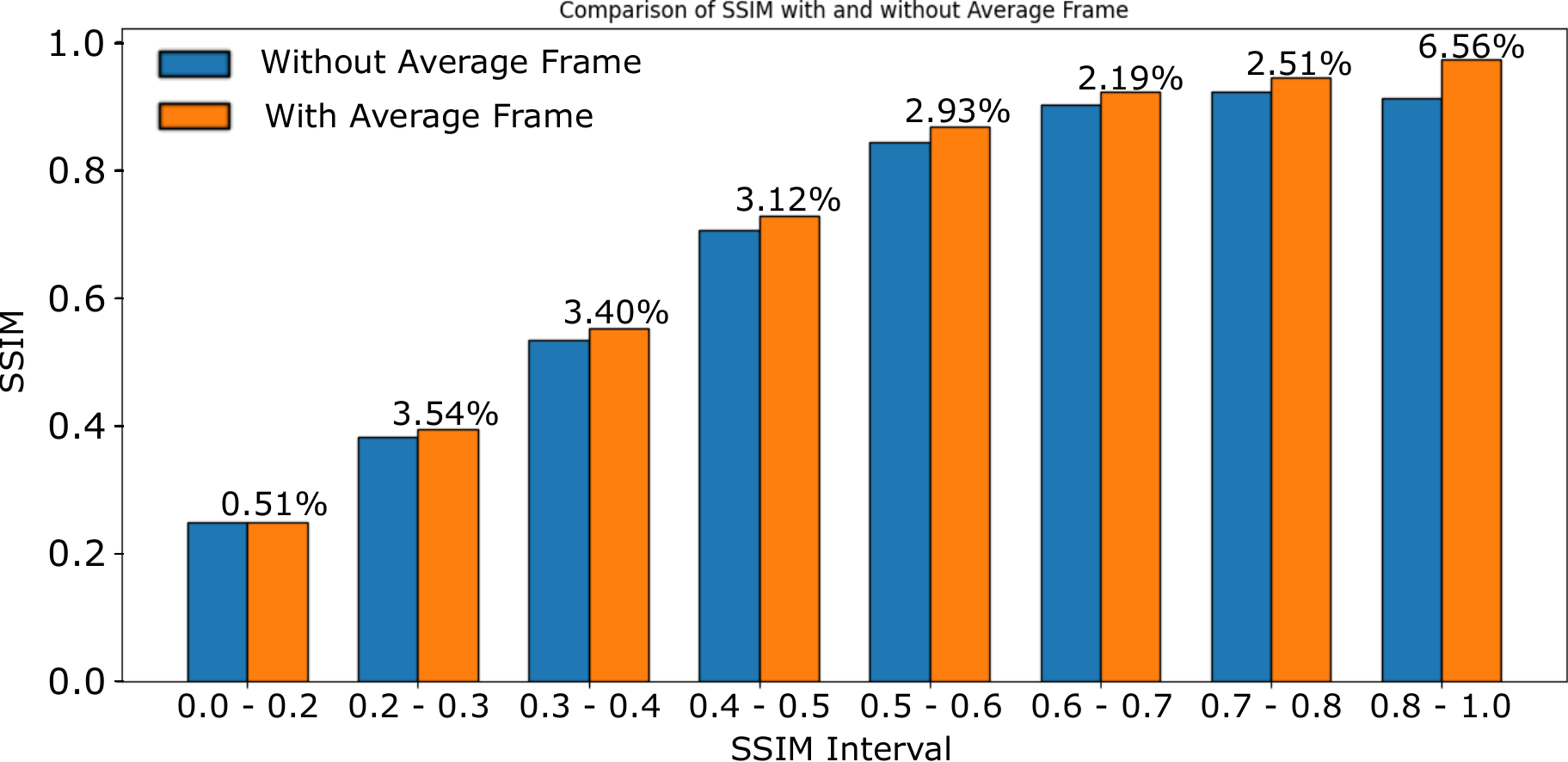}
\end{tabular}
\end{center}
\caption 
{ \label{fig:comparison}
Comparison of SSIM values between the modified TMT model and the modified TMT model incorporating a temporally averaged frame across different SSIM bins. Blue bars represent the modified TMT model, while orange bars correspond to the modified TMT model incorporating the temporally averaged frame. Annotated values indicate the absolute SSIM gain introduced by including the averaged frame.
} 
\end{figure}

In addition to the architectural comparison, Fig.~\ref{fig:comparison} illustrates the impact of including the temporally averaged frame. Across all SSIM bins, the average frame contributes to consistent SSIM improvement. In the lowest SSIM bin (0–0.2), the gain is relatively small (0.0051), suggesting limited utility when distortions are extremely severe. However, as the input images are less affected by atmospheric turbulence (i.e., higher SSIM), the effect becomes more pronounced. The most significant improvement is observed in the 0.8–1 bin, with a gain of 0.0656, indicating that the average frame is particularly effective in refining already plausible reconstructions.

Mid-range bins show steady improvements as well. In the 0.4–0.5 and 0.5–0.6 intervals, the SSIM gains are 0.0312 and 0.0293, respectively. These results suggest that the average frame acts as a strong regularizer, allowing the model to enhance partially restored structures by reinforcing shared spatial features. In bins 0.2–0.3 and 0.3–0.4, moderate gains of 0.0354 and 0.0340 are observed, respectively. This supports the hypothesis that while severely degraded frames provide poor standalone content, the average of multiple frames still captures useful global context.

Notably, the most dramatic effect of the average frame is seen in already high-quality inputs. In the 0.8–1 bin, the Transformer achieves a 0.0656 increase in SSIM. This indicates that the model uses the average frame to fine-tune high-frequency content, correct residual distortions, and reduce minor inconsistencies. The averaging operation suppresses random temporal noise and stabilizes features across frames, giving the model a cleaner reference to anchor the reconstruction.

Several underlying mechanisms may explain this behavior. First, temporal averaging smooths out pixel-level noise and geometric jitter that are inconsistent across frames. This results in a composite frame that better approximates the latent clean image, especially in the presence of mild or moderate distortion. Second, the average frame offers structural guidance that the model can align other frames against during attention-based fusion. This enhances temporal coherence in the output and reduces hallucinations or sharpening artifacts.

Additionally, the average frame may serve a dual role — helping in both low-SSIM and high-SSIM contexts. In severely degraded inputs, it provides missing structural cues that individual frames cannot supply. In high-SSIM settings, it allows the model to focus on correcting subtle pixel-level artifacts that would otherwise be overlooked.

Overall, the results demonstrate that incorporating an average frame significantly enhances reconstruction performance, particularly in cases with moderate to high initial SSIM. While its benefit in extreme distortion is limited, the consistent improvement across bins validates its inclusion in the Transformer pipeline. Combined with the architecture’s inherent attention-based feature extraction, the averaging strategy reinforces the model’s ability to generate sharper, more stable reconstructions across a wide range of turbulence conditions.

\subsubsection{Qualitative Comparison (Visual Assessment)}

This section presents a Qualitative comparison of the restoration performance of all evaluated algorithms across different turbulence regimes. The comparison considers SSIM improvements across stratified bins in order to analyze performance trends as a function of degradation severity.

When the input image has moderate quality (SSIM $\approx$ 0.6), all stabilization methods enable the GEN model to reach a high reconstruction regime with SSIM values close to 0.90. In this regime, performance differences across stabilization strategies become negligible (on the order of $10^{-3}$), and visual distinctions are minimal. This behavior indicates that once a sufficiently stable input representation is achieved, the reconstruction quality becomes generator-dominated, and the specific choice of stabilization method has limited impact, revealing a saturation effect in GEN-based restoration.

\begin{figure}
\begin{center}
\begin{tabular}{c}
\includegraphics[width=16cm]{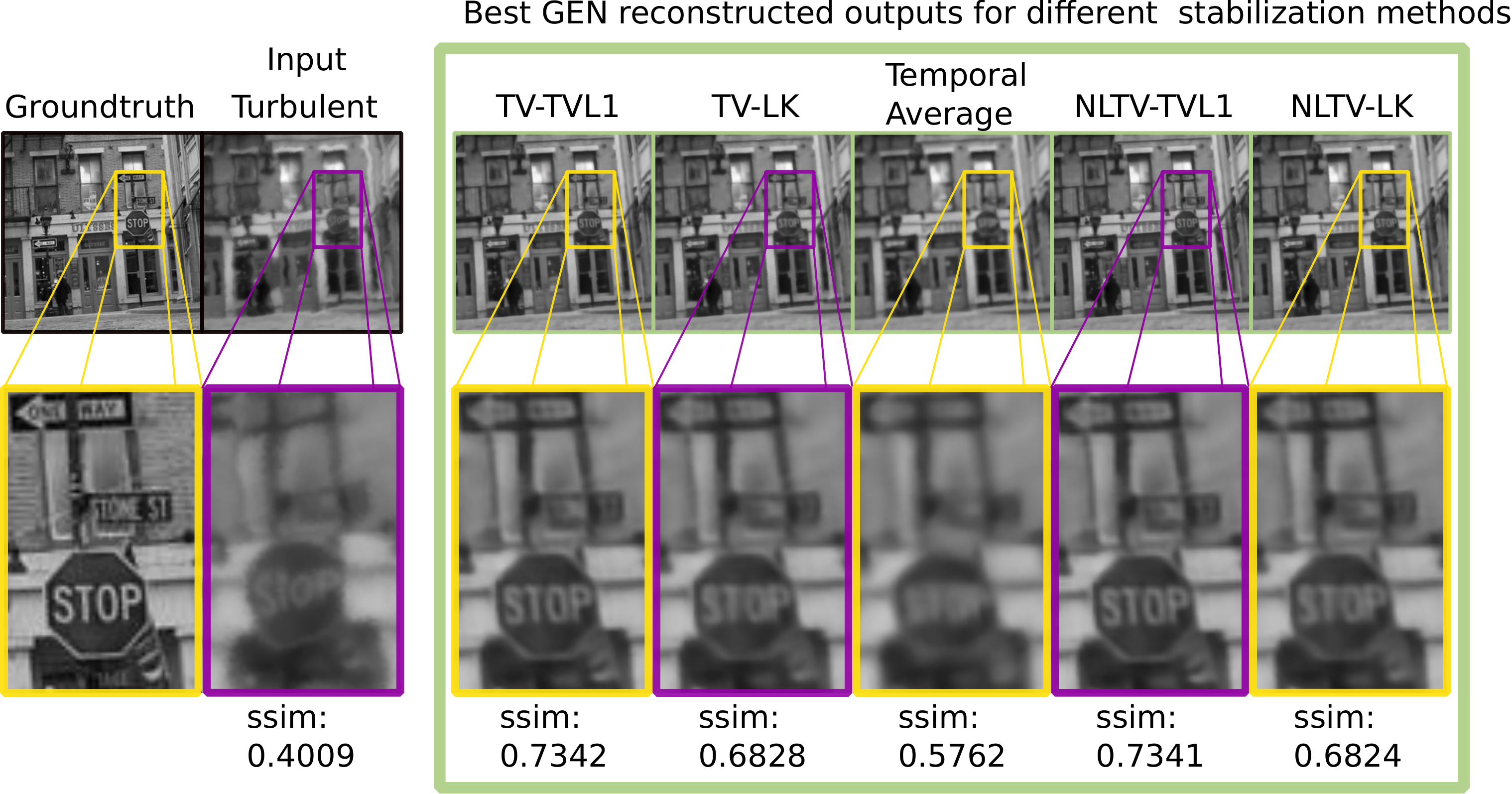}
\end{tabular}
\end{center}
\caption 
{ \label{fig:outputGenExample2}
GEN reconstructed outputs for different stabilization methods starting from a low SSIM input, illustrating the strong dependence of reconstruction quality on stabilization choice in severe turbulence conditions.} 
\end{figure}

Figure~\ref{fig:outputGenExample2} presents GEN only reconstruction results for a more challenging input case, where the turbulence degraded image exhibits a substantially lower initial structural similarity, with an SSIM of approximately 0.40. In this low quality regime, the influence of the stabilization method used to generate the input becomes much more pronounced. While all reconstructions improve upon the turbulent input, the resulting SSIM values vary significantly across stabilization strategies.

In particular, reconstructions obtained using TV-TVL1 and NLTV-TVL1 achieve the highest SSIM values, both exceeding 0.73, and exhibit visibly sharper edges and improved structural coherence in high contrast regions such as the traffic sign and building contours. In contrast, methods such as TV-LK, NLTV-LK and temporal averaging yield more moderate improvements, with SSIM values in the range of 0.57 to 0.68.

These results indicate that, in the presence of severe turbulence and low initial image quality, the choice of stabilization method plays a critical role in enabling effective generator based restoration. Unlike the high SSIM regime illustrated previously, where reconstruction quality saturates and differences between methods become marginal, this example demonstrates that stronger stabilization techniques are necessary to provide inputs that allow the GEN model to recover meaningful structural information. This behavior highlights a transition between two regimes: a stabilization limited regime at low input quality, and a generator dominated regime once a sufficiently stable input representation is achieved.

\begin{figure}
\begin{center}
\begin{tabular}{c}
\includegraphics[width=16cm]{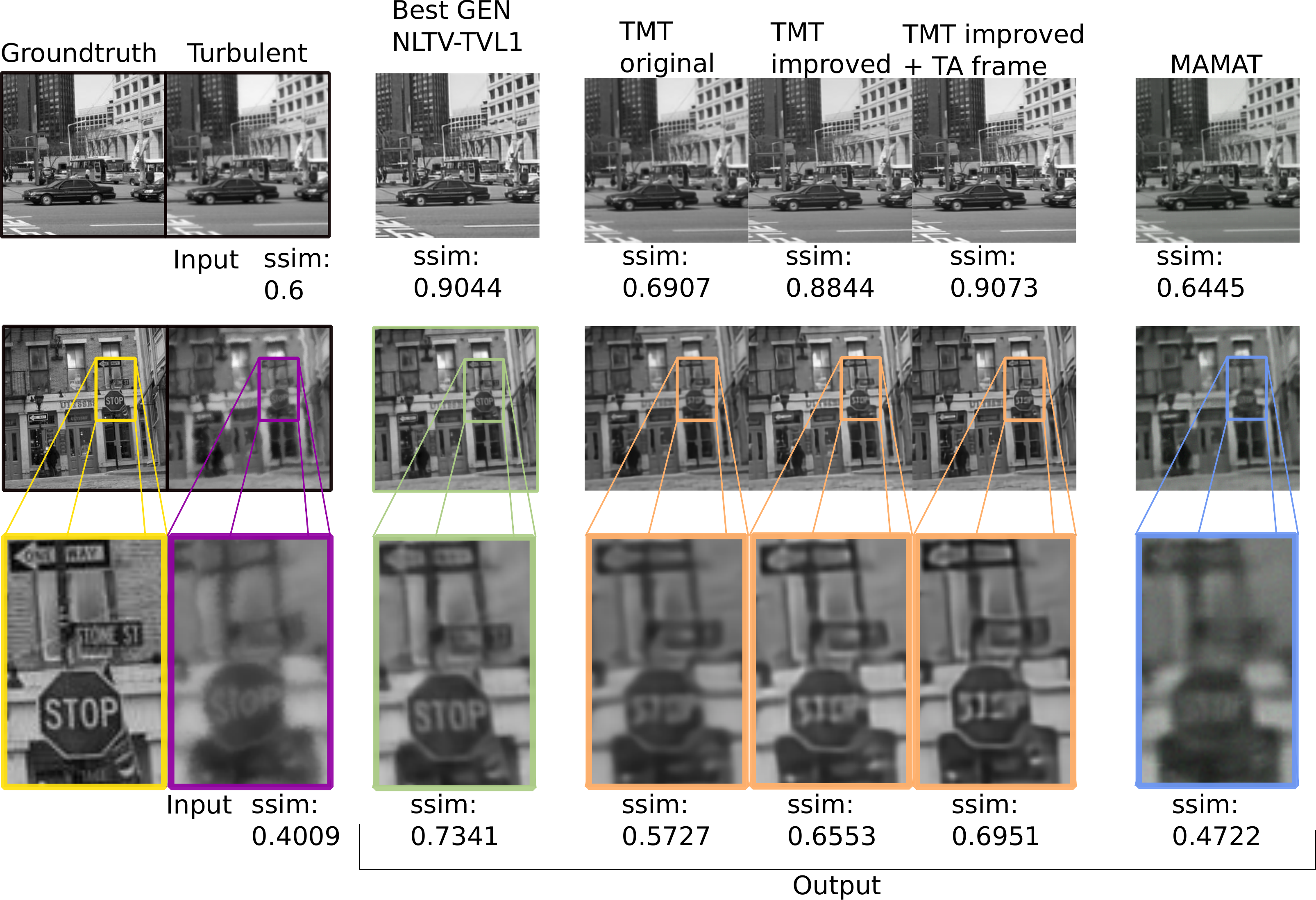}
\end{tabular}
\end{center}
\caption 
{ \label{fig:bestOutputExample}
Comparison of reconstructed outputs from GEN, TMT variants, and MAMAT, illustrating the superior global contrast and edge definition achieved by the Transformer based models despite occasional high SSIM values in GEN reconstructions.} 
\end{figure} 
Figure~\ref{fig:bestOutputExample} presents a qualitative comparison of all evaluated restoration approaches, including the GEN model, the original and modified TMT configurations, and the MAMAT reference method. Overall, the Transformer based model achieved the most visually consistent reconstructions, characterized by higher contrast, better edge definition, and improved preservation of global scene structure. These properties are particularly evident in regions with strong geometric features, such as building edges and vehicle contours.

The GEN reconstruction exhibits a different behavior. While it achieves high SSIM values in this example and provides improved legibility of fine textual details, such as street signs, the output also shows increased smoothing in surrounding regions. This local blur reduces global structural consistency and explains why, despite occasionally higher SSIM values in isolated examples, the GEN model underperforms the Transformer based approaches on average, as shown in the quantitative results summarized in the SSIM bin tables.

The modified TMT model, and in particular the version incorporating the temporally averaged frame, provides the most balanced restoration. It preserves sharp edges and global contrast while maintaining sufficient local detail, leading to consistently higher SSIM scores across a wide range of turbulence conditions. The MAMAT output shows moderate improvement over the turbulent input but remains visually less sharp and structurally coherent than the Transformer based models, consistent with its role as an external reference method.

This comparison highlights an important distinction between local perceptual clarity and global structural fidelity. While generator based approaches may enhance specific features such as text readability, Transformer based models offer more robust and consistent restoration by jointly modeling spatial and temporal information, making them better suited for general atmospheric turbulence mitigation.

\subsubsection{Global Performance Comparison Across Models}

\begin{figure}
\begin{center}
\begin{tabular}{c}
\includegraphics[width=16.3cm]{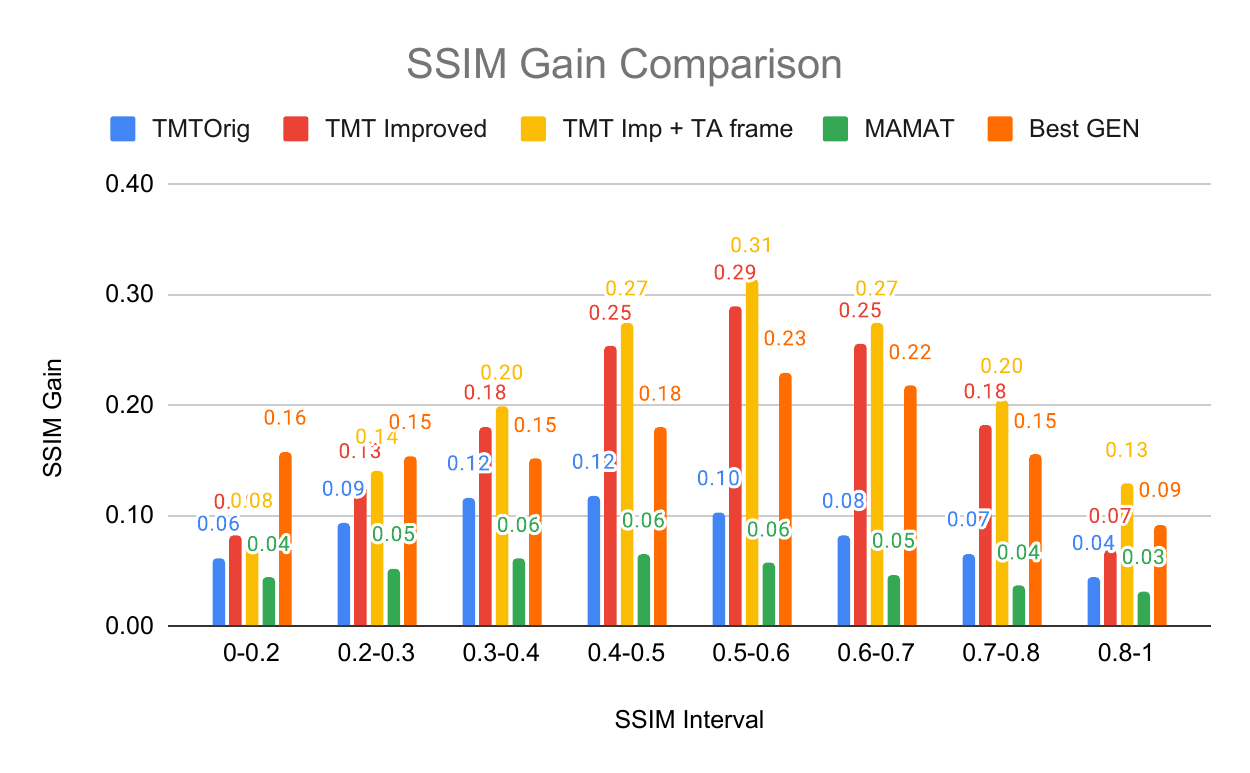}
\end{tabular}
\end{center}
\caption 
{ \label{fig:comparisonFianl}
Comparison of SSIM gains for all algorithms across different SSIM regimes}, highlighting the superior and more consistent performance of the improved Transformer based models.
\end{figure} 
Figure~\ref{fig:comparisonFianl} provides a global comparison of SSIM percentage improvement across all evaluated algorithms and turbulence regimes. The comparison across all algorithms reveals a nuanced performance trade off that depends on the severity of the input degradation. In the most challenging turbulence regimes, corresponding to the lowest SSIM bins $(0\text{--}0.2]$ and $(0.2\text{--}0.3]$, the GEN model achieves the highest SSIM improvement among all evaluated approaches. These bins represent the most severely degraded inputs, indicating that generator based restoration can be particularly effective in recovering structural similarity when turbulence effects are extreme.

As the input quality improves and turbulence severity decreases, the Transformer based models progressively outperform the GEN approach. In the low to mid SSIM regimes, the improved TMT configurations, and especially the version incorporating a temporally averaged frame, achieve larger and more consistent SSIM gains. This behavior reflects the advantage of explicitly modeling temporal information and enforcing structural consistency across frames, which becomes increasingly important as geometric distortions become less dominant than residual blur and fine scale artifacts.

The GEN model shows competitive performance in several bins and, in some cases, achieves higher local SSIM improvements than the original TMT. This behavior is consistent with the qualitative examples, where GEN reconstructions often enhance the legibility of fine details such as text. However, the overall SSIM improvements achieved by GEN remain lower than those obtained by the modified TMT models when averaged across the dataset, reflecting increased smoothing and reduced global structural consistency in the reconstructed outputs.

The original TMT and the MAMAT reference method show more modest and stable improvements across bins. While both approaches consistently outperform the turbulence degraded inputs, they do not match the gains achieved by the modified Transformer configurations. Overall, this comparison confirms that combining multi frame attention with improved training strategies and an additional averaged reference frame yields the most effective and consistent restoration performance, making the enhanced TMT model the best performing approach in terms of SSIM across the full range of turbulence conditions.



\subsection*{Disclosures}
Large Language Models (LLMs) were employed solely to refine the clarity, grammar, and fluency of the manuscript. Specifically, OpenAI’s ChatGPT was used during the final stages of writing to enhance the precision and readability of the text. No content generation, experimental design, data analysis, or figure creation was performed by AI tools. All scientific contributions, methodology, and results are entirely the product of the authors. The use of ChatGPT was limited to editorial assistance to improve the quality of exposition.

The prompts used were restricted to language-editing tasks and did not involve generating scientific content. Representative examples include: "Improve the grammar and clarity of the following paragraph while preserving its original meaning," "Rewrite this sentence in a more formal academic tone," and "Check for grammatical errors and enhance readability without altering technical content." These prompts were used exclusively for editorial refinement.

\subsection* {Code and Data Availability} 
The SOTIS dataset used in this study is publicly available at \url{https://zenodo.org/records/18500766}.
The code used to produce the results in this study is publicly available at the following GitHub repository:

\url{https://github.com/sebastiancarrillo/gan-vs-transformerSOTIS}

The repository includes:
\begin{itemize}
    \item Python scripts implementing the Transformer-based models for atmospheric turbulence mitigation, including training, evaluation, and preprocessing routines.
    \item A Jupyter notebook demonstrating the GAN-based approach, including generation, evaluation, and visual comparison.
\end{itemize}

All experimental results in the manuscript can be reproduced using the provided code and access to the SOTIS dataset.


\subsection* {Conflicts of interest}
The authors declare that there are no financial interests, commercial affiliations, or other potential conflicts of interest that could have influenced the objectivity of this research or the writing of this paper.

\subsection* {Acknowledgments}
This work was funded in part by the Air Force Office of Scientific Research, grant FA9550-21-1-0275 and by the U.S. National Science Foundation (Award IIS 2324008).



\appendix    


\bibliography{report}   
\bibliographystyle{spiejour}   

\vspace{2ex}\noindent\textbf{Juan S. Carrillo} is a PhD student in the Joint Doctoral Program in Computational Science at San Diego State University and the University of California, Irvine. His research focuses on deep learning, computer vision, and image processing, with applications in turbulence mitigation and medical imaging. He has co-authored four peer-reviewed publications, including work on glaucoma detection using neural networks. His current work explores transformer-based architectures for real-world image restoration tasks.


\vspace{2ex}\noindent\textbf{Jérôme Gilles} received an M.S. in signal processing (1999) and a Ph.D. in mathematics (2006) under Prof. Yves Meyer from ENS Cachan, France, and completed his Habilitation in 2012 with Prof. Jean-Michel Morel. He worked for the French Ministry of Defense (2001–2010) and UCLA (2010–2014), joined SDSU in 2014, and became full professor in 2022. His research focuses on signal/image processing, wavelets, and applied mathematics.

\vspace{1ex}

\listoffigures
\listoftables

\end{spacing}
\end{document}